\documentclass[aps,pra,floatfix,twocolumn,superscriptaddress,nofootinbib,twocolumn]{revtex4-1}
\usepackage{xr}

\usepackage{physics}
\usepackage{amsmath}
\usepackage{amsthm}
\usepackage{amsfonts}
\usepackage{amssymb}
\usepackage{textcomp}
\usepackage{graphicx,float}
\usepackage{bm}
\usepackage{color}
\usepackage{verbatim}
\usepackage{graphicx,color}
\usepackage{hyperref}
\usepackage{enumitem}

\hypersetup{
colorlinks=true,
linkcolor    = blue,
citecolor   = blue,
    pdfborder={0 0 0},
}

\newcommand{\cov}{\nabla}
\newcommand{\p}{\partial}

\newcommand{\la}{\langle}
\newcommand{\ra}{\rangle}

\newcommand{\rar}{\rightarrow}

\newcommand{\half}{{1\over2}}
\newcommand{\OO}{{\cal O}}
\newcommand{\bp}{\mathbf{p}}

\theoremstyle{definition}

\definecolor{orange}{rgb}{1,0.5,0}
\definecolor{col1}{RGB}{153, 52, 121}

\bibpunct{[}{]}{,}{n}{}{}

\begin{document}
\title{
Anomalous attenuation of plasmons in strange metals and holography
}

\author{Aurelio Romero-Berm\'udez}
\author{Alexander Krikun}
\author{Koenraad Schalm}
\affiliation{Instituut-Lorentz, $\Delta$ITP, Universiteit Leiden, P.O. Box 9506, 2300 RA Leiden, The Netherlands}
\author{Jan Zaanen}
\affiliation{Instituut-Lorentz, $\Delta$ITP, Universiteit Leiden, P.O. Box 9506, 2300 RA Leiden, The Netherlands}
\affiliation{Department of Physics, Stanford University, Stanford CA 94305, USA}

\begin{abstract}

The plasmon is a ubiquitous collective mode in charged liquids. Due to the long-range Coulomb interaction, the massless {zero} sound mode of the neutral system
acquires a finite plasmon frequency in the long-wavelength limit. In the zero-temperature state of conventional metals -- the Fermi liquid -- the plasmon lives infinitely
long at long wavelength when the system is (effectively) translationally invariant. 
In contrast, we will show that in strongly entangled strange metals 
the protection of zero sound fails at finite frequency and plasmons are always short lived regardless of their wavelength. 
Computing the explicit plasmon response in holographic strange metals as an example, we show that decay into the quantum critical continuum replaces 
Landau damping and this happens for any wavelength.

\end{abstract}
\maketitle

\section{Introduction} 

The plasmon is a ubiquitous propagating mode in electromagnetically charged systems. It is governed by a simple universal principle. Any system, that in the absence of electromagnetic interactions, carries a propagating gapless density mode with dispersion $\omega = v_s \mathbf{|p|}$ will, upon switching on 
the static Coulomb interaction, see this dispersion change to $\omega  = \sqrt{ \omega^2_p + v_s \mathbf{p}^2 }$ with $\omega^2_p=\rho^2/(\epsilon+p)$ the plasma frequency;
 here $\rho$ is the (charge) density, $\epsilon$ is the~energy density, and $p$ is the pressure. In nonrelativistic systems this becomes $\omega^2_p = n e^2 / m$
with $e$ and $m$ the charge and the mass of the charged particles at a density $n$. The plasmon's ubiquity ranges from crystals formed from charged particles
in a homogeneous
background, where the longitudinal phonon becomes the plasmon \cite{2dcrystals},  
to superconductors. {Here the plasmon originates in the Nambu-Goldstone boson through the Anderson-Higgs mechanism and takes the role of the longitudinal photon polarization.}\footnote{In relativistic systems the longitudinal and transversal polarizations combine in order to form massive vector bosons with mass equal to the plasmon frequency. In nonrelativistic
superconductors characterized by a material velocity $v_F \ll c$ the mass for the transversal vector polarizations, which sets the London penetration depth, is a factor $c/v_F$ smaller than $\omega_p$.}        

Here we focus on the plasmons found in the zero-temperature metallic state formed by electrons in solids. Normal metals are described by the Fermi-liquid (FL) theory coupled to electromagnetism and the 
explanation of plasmons is part of its success. 
The zero-temperature uncharged Fermi liquid with repulsive $s$-wave interactions (the Landau parameter $F^s_0 >0$) is characterized by a propagating longitudinal density mode with dispersion $\omega = v_s\mathbf{|p|}$: the zero sound. Similar to the usual "first sound" of classical fluids, it is hydrodynamically 
protected in the sense that at low energies and long wavelengths momentum conservation and particle number conservation inhibit its decay. 
Upon including the long-range Coulomb interaction, this zero sound is promoted to the plasmon which requires a finite frequency $\omega_p$ to excite.

\begin{figure*}[t!]
	\center
	\includegraphics[width=0.43 \linewidth]{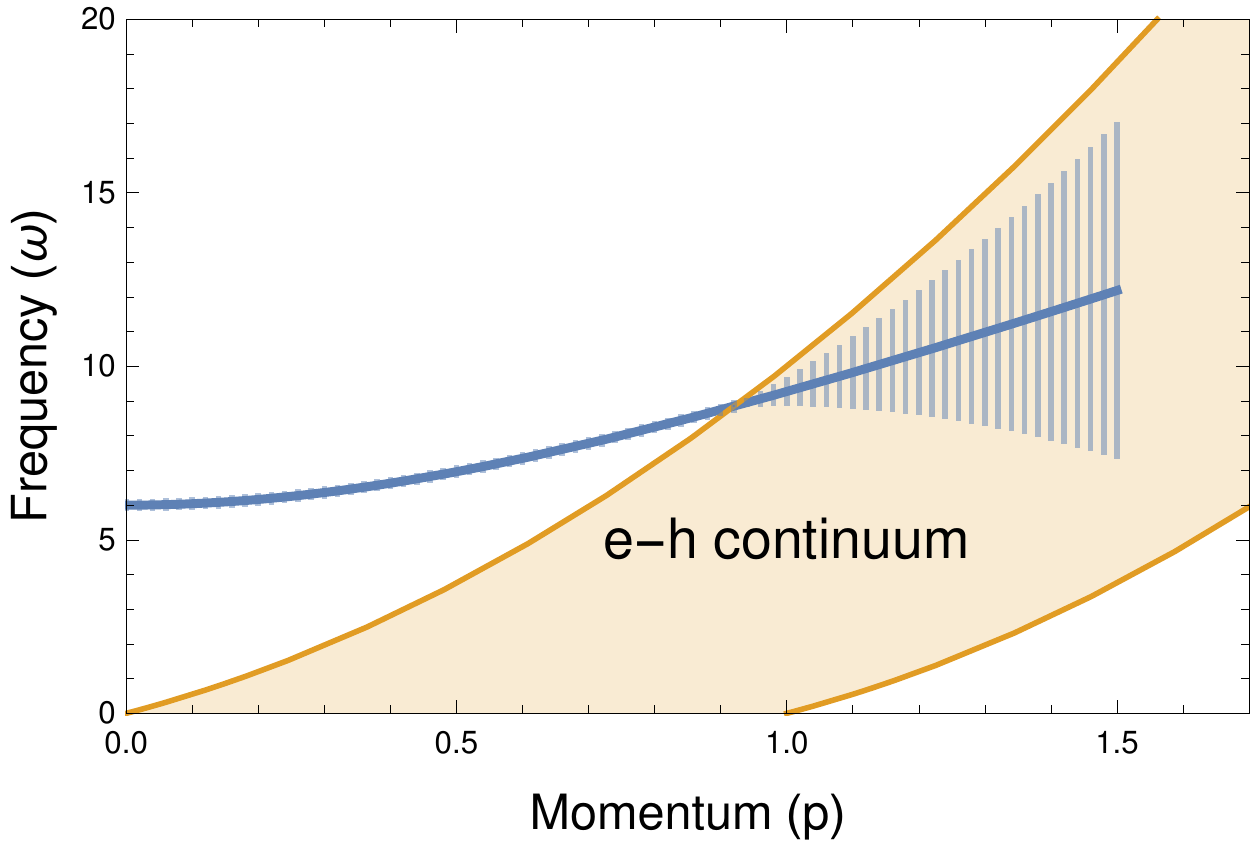}
	\includegraphics[width=0.43 \linewidth]{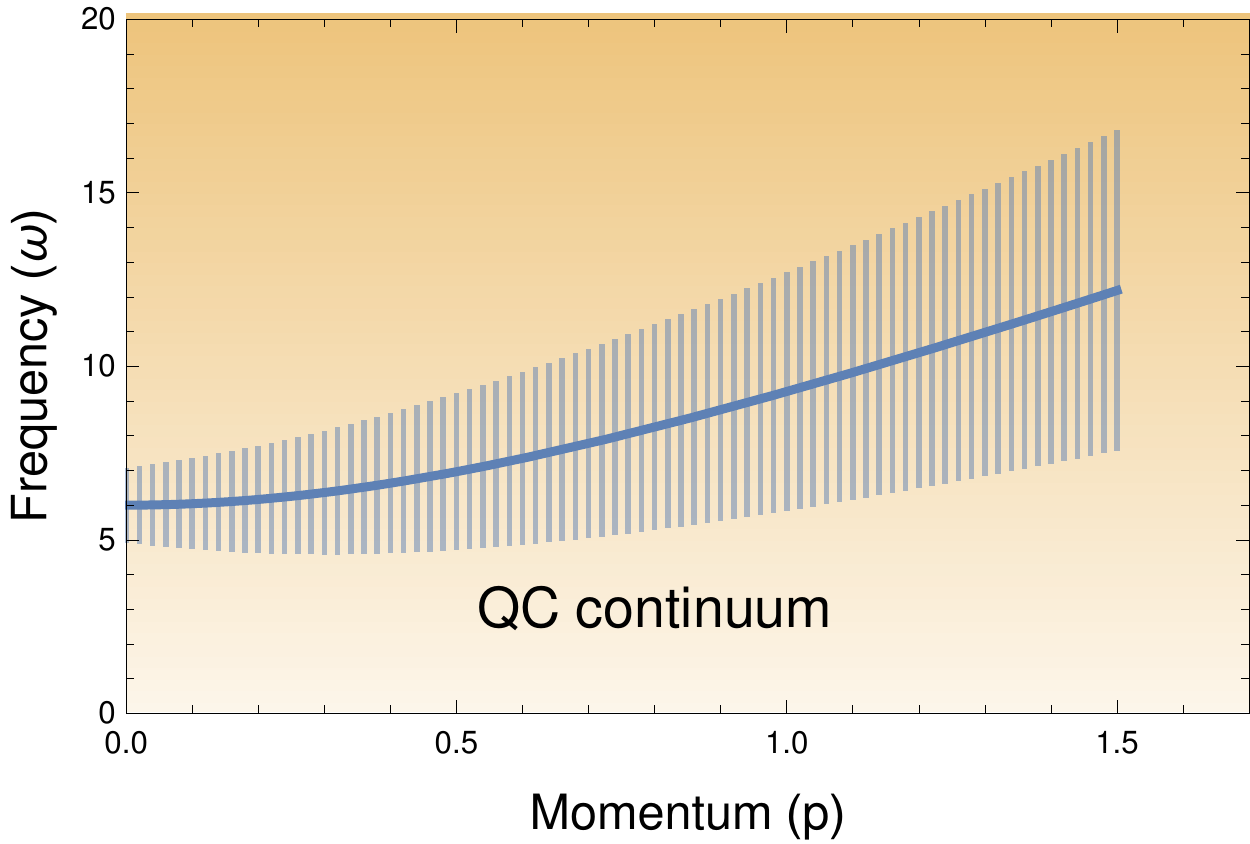}
	\vspace{-2mm}\caption{\label{fig:cont} In conventional Fermi liquids the width of the plasmon (blue lines) at large momenta is due to decay into the particle-hole (Lindhard) continuum (left). In strange metals there is a quantum critical continuum at all frequencies and momenta, and a plasmon in a strange metal can therefore decay at any momentum (right). 
	}
\end{figure*}

At this finite frequency, the universality of hydrodynamical-like protection no longer applies {and one would expect the plasmon to decay. Nevertheless, the Fermi-liquid plasmon is stable. This is well understood thanks to the fact that the Fermi liquid} renormalizes into a system of noninteracting quasiparticles. Using this microscopic description one can compute that,  besides the propagating plasmon, the Fermi-liquid can also react to electromagnetic probes by ``incoherent'' particle-hole excitations. The phase space for these excitations is limited by energy and momentum conservation. 
This phase space -- the Lindhard continuum -- shrinks to { a small region near} zero frequency
in the long-wavelength limit (see Section~\ref{sec:zero_sound}). Therefore at low momenta the plasmon is completely isolated from these excitations{; it has no channels to decay} and inherits thereby the hydrodynamical protection of the zero sound.
At larger momenta the plasmon dispersion will cross the upper bound of the continuum, at which point it will rapidly decay in uncorrelated particle-hole excitations: 
the classic Landau damping (see Fig.\,\ref{fig:cont}). 

In any real metal, translational symmetry is broken by the atomic lattice and momentum is not strictly conserved. Hydrodynamical 
protection in the strictest sense therefore does not apply{, neither does it apply for zero sound or its inheritance by the plasmon.} However, in weakly interacting Fermi liquids as realized in simple metals, the resulting Umklapp scattering 
becomes noticeable in the zero-momentum charge response only upon reaching the energy of interband transitions.  
When $\omega_p$ is not coincident with this energy, the lattice continues to be irrelevant and long-wavelength plasmons continue to be long lived. 
This physics is at the basis of a flourishing  engineering  field that makes use of the beneficial properties of the plasmon as a ``material photon'': plasmonics. 

However, these well established wisdoms are violated in a spectacular fashion in the non-Fermi-liquid strange metals realized in, e.g., the cuprate high-T$_c$ superconductors.
In the $\mathbf{p} \rightarrow 0$ limit the plasmons can be studied via optical conductivity and are found to be almost overdamped: their inverse life time is of the order of the plasmon frequency \cite{BozovicOpt,SlakeyOpt}. 
This is surely not due to an energy independent elastic scattering since the momentum relaxation rate observed in the dc conductivity at low temperatures is smaller by orders of magnitude than the observed plasmon attenuation. 
The charge density response at finite $\mathbf{p}$ has been measured as well with a limited 
energy resolution using high-energy transmission energy electron-loss spectroscopy. These data reveal 
a dispersing  plasmon  becoming completely overdamped at a momentum 
of approximately $1/3$ of the Brillouin zone \cite{nucker1989plasmons,nucker1991long,fink1994}. Very recently the data obtained by low-energy reflection electron-loss spectroscopy were presented \cite{mitrano2018anomalous}. Although these data suggest that the plasmon disappears even faster for increasing momentum, these also appear to reveal directly the nature of the incoherent excitations. Instead of the Lindhard continuum it was found that the spectrum is independent of momentum and energy up to a high-energy cut-off at $\simeq 1$ eV. 

It is clear that the atomic lattice effects and the Umklapp scattering are less ``hidden" in these strongly interacting electron systems as compared to the simple metals. 
This has already been seen in exact diagonalization studies on small Hubbard model clusters \cite{tohyama2005exact} and in  recent quantum Monte Carlo results \cite{huang2018strange}. 
For strange metal models described by a holographic framework, which we will be considering here, it is also established that lattice effects significantly modify the finite frequency response in optical conductivity \cite{horowitz2012optical,donos2015thermoelectric,langley2015absence}. Undoubtedly, these effects have to be dealt with in order to address what is going on in experiment. 

This we will not {pursue.} We will address instead a matter of principle of such a generality that it may well play a crucial role in any attempt to explain experiment. 
Consider a translationally invariant system, eliminating the complicating lattice effects. 
What is the fate of the plasmon in a metal that does {\em not} 
renormalize to the free fermion fixed point controlling the Fermi liquid? 

We use the holographic strange metals as a model system which we claim is in the present context fully representative {for such non-Fermi liquids}. 
A defining characteristic of these systems is that their charge response is characterized by two "sectors": the zero sound as in the neutral Fermi liquid and, in addition, a "quantum critical" (QC) continuum \cite{Zaanen:2015oix,Hartnoll:2016apf,Zaanen:2018edk}. 
{The additional significance of this quantum critical sector is that it cannot be understood in terms of particle-like excitations and it therefore has a unique imprint in transport and collective responses. This sector} should not be confused, however, with the fluctuations associated with a quantum critical point. The holographic strange metal shows scaling behaviors characterized by hyperscaling violation ($\theta$) and dynamical critical ($z$) exponents, which are set by the deep infrared quantum critical sector of the dual model but which  are unrelated to the familiar ``bosonic'' quantum criticality.

\onecolumngrid
\twocolumngrid

This very presence of the QC sector in holographic models{, seemingly unrelated to anything that is familiar in conventional condensed matter, has been a puzzle. Most of the holographic strange metal studies are focused on the zero-momentum collective observables  (rare exceptions include \cite{Hartnoll:2012wm,Anantua2013,Blake:2014lva}). Collective responses in the Fermi liquid at zero momentum, on the other hand, lack any kind of a ``second sector'' contribution. However, this situation changes once one turns to finite momentum. Then an algebraic second sector arises in Fermi liquid as well: the Lindhard continuum. Clearly, it is very different from the scaling infrared continuum of the strange metal, as we will discuss in more detail below.} However, effectively, the holographic
strange metals may be indeed viewed as ``strongly interacting'' (in the critical theory sense) generalizations of the Fermi liquid, {with two-sector collective responses} where the QC sector supplants the Lindhard continuum.

From this perspective, the vanishing of the Lindhard continuum at zero momentum is a {\em singular} feature of the free fermion fixed point. In a non-Fermi liquid an incoherent continuum should be present at {any finite frequency and momentum}. The ramification is that, as we show below, {\em even in the perfect Galilean continuum  the long-wavelength limit plasmon will decay in finite time} (see Fig.\,\ref{fig:cont}).
The infinitely long lifetime of the Fermi-liquid plasmon is not rooted in general principle, but instead in the singularly special nature of its free particle IR fixed point.

In the remainder of this paper we will provide the evidence for this case. In  Sec.\ref{sec:zero_sound} we will remind the reader of the results of electromagnetic linear response
theory in metals, tying together the electrical conductivity with the charge density response and highlighting the universality of the plasmon. We will also specify how this can be married with the 
nominally neutral matter computed by holography.     
In Sec.\ref{sec:holographic_model} we introduce the holographic models for strange metals and show how the quantum critical sector is encoded in the geometry. We also 
 review how their zero sound response is encoded in quasi-normal mode (QNM) fluctuations of this dynamical space-time geometry. 
In Sec.\ref{sec:plasmon} we show how the {long-range} Coulomb interaction can be encoded in the holographic dictionary as a so-called double trace deformation{, a point also made in the recent article \cite{Mauri:2018pzq}.} This allows us to 
compute the corresponding charged density-density response function and study the corresponding modified quasi-normal mode spectrum.
Our results are consistent with earlier 
studies of the plasmon in holographic strange metals where the charge response is computed in a different manner
 \cite{gran2018exotic,aronsson2017holographic,aronsson2018plasmons}. This central section also contains the main results: {as a consequence of the omnipresent QC sector into which the holographic strange metal plasmon can decay,} the width of the plasmon is significantly enhanced 
and finite even at zero momentum. We devote Sec.\ref{sec:tuned} to the study of the dependence of the features of the plasmon mode on the 
characteristics of the quantum critical sector by tuning its scaling dimensions. 
The surprise is that the plasmon width at zero momentum is quite insensitive to these vast changes in the deep IR. 
Appendix \ref{sec:Ax} is devoted to how to go beyond the static Coulomb interaction and include dynamical electromagnetism.  Appendix \ref{app:dressings} generalizes our three-dimensional (3D) results to an effective two-dimensional (2D) Coulomb interaction 
$V_\bp = e^{-\lambda|\bp|}/|\bp|$.

\vspace{-.05in}

\section{Electromagnetic linear response theory: the Plasmon as dressed zero sound.}
\label{sec:zero_sound}

We are envisioning an electrically charged liquid. Let us first collect central results from the standard theory describing the linear response of such a system to external 
electromagnetic fields \cite{mahan2013many,Altland:2006si}.
The central quantity enumerating the   response of the charged medium is the {\em polarization propagator} $\Pi (\omega, \bp)$. To make contact
with condensed matter physics we consider nonrelativistic systems characterized by a material velocity scale that is much smaller than the velocity of light.  In the transversal 
channel $\Pi(\omega, \bp)$ is just the photon self-energy associated with the interaction with  matter. Our focus is here entirely on the longitudinal channel, which encodes the response
to longitudinal electric fields. It is captured by the dielectric function, which relates the displacement to the electrical field $\vec{D} ( \omega, \bp) = \varepsilon ( \omega, \bp) \vec{E}$ (assuming an isotropic medium). The dielectric function $\varepsilon (\omega, \bp)$,  the frequency and momentum dependent conductivity $\sigma (\omega, \bp)$ and the charge density response function $\chi (\omega, \bp)$ are determined in terms of the polarization propagator and Coulomb interaction $V_{\bp} = e^2/ \bp^2$ as
\begin{align} 
\sigma (\omega, \bp) & =   i \frac{\omega}{\bp^2} \Pi (\omega, \bp), \nonumber \\
\varepsilon ( \omega, \bp) & =  1 - V_{\bp} \Pi (\omega, \bp), \nonumber \\
\chi (\omega, \bp) & =  \frac{\Pi (\omega, \bp)}{ 1 - V_{\bp} \Pi (\omega, \bp)}.
\label{EMlinresp}
\end{align}
One notices that the optical conductivity is most closely related to $\Pi(\omega,\bp)$; this is a consequence of the continuity
equation associated with the conservation of electrical charge that links the density and longitudinal currents through  $\partial_t n  + \vec{\nabla} \cdot \vec{J} =0$.
The practical difficulty with the optical conductivity is that it can only be measured at $\bp \rightarrow 0$ because of the smallness of the material velocity compared to the velocity of light. 
There is much to learn at finite momenta, but this is only accessible by mobilizing electrons with their small wavelengths through electron loss spectroscopy; { these measure the loss function $\text{Im}\frac{1}{\varepsilon ( \omega, \bp)}$ or the charge susceptibility $\chi (\omega, \bp)$ directly}.

In a continuum charged Fermi liquid with a repulsive $F^0_s$ Landau parameter the polarization propagator can be parametrized at zero temperature, and not too large momenta 
as
\cite{abrikosov2012methods}
\begin{equation}
\label{equ:zero_sound}
{\Pi(\omega, \bp)} \approx \frac{\bp^2 A}{\omega^2 - (v_s \bp)^2 + i \omega \Gamma(\bp) + \Gamma (\bp)^2/4} + \bp^2\Xi(\omega, \bp).
\end{equation}
This {low momentum response of the polarization operator} $\Pi( \omega, \bp)$ has the same structure as the (particle number) density response $\chi^0(\omega, \bp) = \langle n(\omega,\bp)n(-\omega, -\bp))\rangle$ of a neutral Fermi liquid such as  $^3$He. It is characterized by a zero sound mode with 
velocity $v_s$ and a hydrodynamical-like damping rate  $\Gamma(\bp) = {\cal D} \bp^2$ where ${\cal D}$ is the diffusivity. In the nonrelativistic Fermi liquid its pole strength equals
$A = \langle n\rangle / m$ ($\langle n\rangle$ and $m$ being the carrier density and electron mass, respectively). 
In addition, it contains the incoherent second sector denoted with $\Xi(\omega, \bp)$. This is the Lindhard continuum. In the 3+1 dimensional Fermi gas it has the well known branch cut form, 
\begin{equation}
\Xi_{FG, d=3} (\omega, \bp ) \sim 1 + \frac{\omega}{ 2 \bp v_F} \text{ln} \left| \frac{ \omega - \bp v_F}{\omega + \bp v_F} \right|.
\label{3DLindhardt}
\end{equation}
This continuum is bounded from above {by energy and momentum conservation for a quasi-particle-hole pair near the Fermi surface:} $\omega_{max} (\bp ) = \frac{p_F |\bp| }{m} +  \frac{\bp^2}{2m}$. The bounding value $\omega_{max}$ goes to zero in the long-wavelength limit. 
One observes that the continuum obeys a $z=1$ dynamical critical scaling while at finite momentum its imaginary part increases linearly at low frequency in $d \ge 2$ space dimensions (see, e.g., \cite{Mihaila2011}). 

It follows from Eq.\,\eqref{EMlinresp} that for small $\bp$ the conductivity behaves as
\begin{equation}
\sigma (\omega, \bp){=}  \frac{i \omega A}{\omega^2 {-} (v_s \bp)^2 {+} i \omega \Gamma(\bp) {+} {1\over4}\Gamma (\bp)^2} {+} i \omega \Xi (\omega, \bp),
\label{conducgen}
\end{equation}
{
or in the strict $\bf{p}\rightarrow 0$ limit
\begin{align}
  \label{eq:5cc}
  \sigma (\omega) =  \frac{i A}{\omega} + i \omega \Xi (\omega, \bp=0).
\end{align}
The second term is well known in holography as the ``incoherent conductivity''  \cite{Hartnoll:2007ih,Blake:2013bqa,Davison:2014lua,Davison:2015bea,Davison:2015taa}, arising from the quantum critical sector of the model.
In a Fermi-liquid this contribution vanishes at $\bp=0$, but 
would be seen at finite momentum, being associated in this case with the Lindhard continuum.
However, because $\Xi(\omega,\bp)$ vanishes at $\bp = 0$,} all that remains in the Fermi-liquid is the infinitely long-lived sound showing up in the conductivity as the  diamagnetic delta function at zero frequency  $\sim i A / \omega$. It is a {peculiarity} of the free fermion fixed point that the second sector contribution disappears in the long-wavelength limit.

To finalize {our brief review of electromagnetic response in charged matter}, it also immediately follows from Eq.\,\eqref{EMlinresp} that the charge susceptibility $\chi(\omega,\bp)$ (or the loss function) has a pole when $\Pi (\omega, \bp) = 1/V_\bp$. Substituting in Eq.\,\eqref{equ:zero_sound} one finds a propagating {gapped} excitation with dispersion $\omega = \sqrt { (v_s \bp)^2 + \omega_p^2 }$, damped by $\Gamma$.  {The zero sound gets ``promoted'' to the plasmon when one observes the density response.} {Eventually the plasmon } becomes overdamped  when it ``dives'' in the Lindhard continuum at large momenta: the conventional Landau damping. 

{We wish to address the electromagnetic response in strange metals described by holography in the next section, but to do so} we face a problem of principle: the standard holographic strange metals are electrically {\em neutral} fluids. The intrinsic strong interactions, which form a holographic strange metal, are in principle not of electromagnetic origin.  
{%
There is, however, a loophole. 
In the treatment presented above the {Coulomb interactions were separated into two parts. The long-range Coulomb force is accounted for perturbatively via the factors $V_\bp$. 
On the other hand the short-ranged part of the screened Coulomb interactions gives rise} to the polarization propagator $\Pi(\omega,\bp)$ of the charged electron system. A typical representative of this kind is the Hubbard $U$ potential that may well be of crucial importance for the emergence of the real strange metals. 
As we noted below Eq.\,\eqref{equ:zero_sound},
the polarization propagator in charged electron Fermi liquids, however, shares the same zero sound structure as the density response in neutral Fermi liquids such as ${}^3$He. Computing the density response in a neutral Fermi system can therefore be used as an analogy for computing the short-range electromagnetic response in a charged Fermi-system.
Similarly, we can thus} consider the strong intrinsic interactions of the neutral holographic model to correspond to short range Coulomb forces and regard the holographic strange metals as a model system that captures generic features of the charged strange metals. 
Subsequently the long-range part of the Coulomb interaction can be regarded as weakly coupled and treated in exactly the same fashion as above.

To {make this more precise,} let us specify some formalism. The model system should be characterized by a conserved current  
$J^{\mu}$ with $n(p) \equiv J^0(p), \ p=(\omega, \bp)$. This current is  externally coupled to dynamical electromagnetism with a strength~$e$ {--- in relativistic notation,
\begin{align}
  \label{eq:6}
  \partial_{\mu} F^{\mu\nu} = e J^{\nu}.
\end{align}
 This equation of motion together with the dynamics of the strange metal is conveniently encoded in the action for the electromagnetic potential $A_{\mu}$}
\begin{align}
  \label{eq:1b}
  S {=} S_{SM} {+} \int \!\!\text{d}^3x\text{d}t\left[ eA_{\mu}J^{\mu} {+}\frac{1}{2} A_{\mu}\big(\delta^{\mu}_{\sigma}\partial^{\nu}\partial_{\nu}{-}\partial_{\sigma}\partial^{\mu}\big)A^{\sigma}\right].
\end{align}
Fixing a Lorentz gauge $\partial_{\mu}A^{\mu}=0$,  the electromagnetic potential can be integrated out. In the absence of external photon sources the effective action becomes
\begin{align}
  \label{eq:1}
  S = S_{SM} + \int \!\!\frac{\text{d}^4p}{(2\pi)^4}~ \frac{e^2}{2} J^{\mu} \frac{\eta_{\mu\nu}}{p^2}J^{\nu}  .
\end{align}
For nonrelativistic condensed matter systems characterized by  velocities $v_F \ll c$ and $\eta_{00} \gg \eta_{ii}$ the electromagnetic interaction reduces to the static 3D
Coulomb interaction, 
\begin{align}
  \label{eq:2}
  S = S_{SM} - \int \!\!\frac{\text{d}^3\bp\text{d}\omega}{(2\pi)^3}~ \frac{1}{2} n(-\omega,-\bp) V_{\bp}n(\omega,\bp).
\end{align}

\begin{figure}[t]
	\center
	\includegraphics[width=0.8\linewidth]{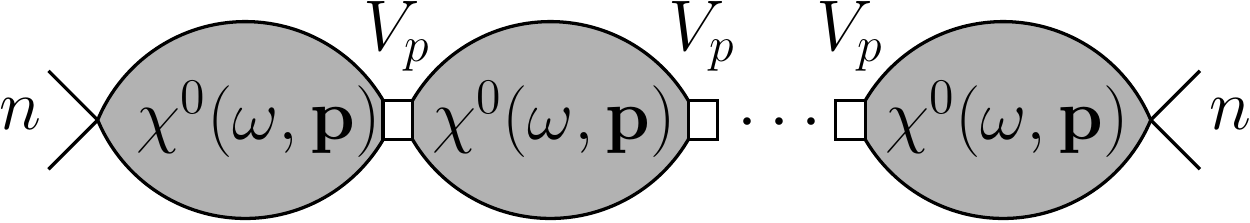}
	\caption{\label{fig:RPA} Dyson resummation  for the ``dressed'' charge density two-point correlation function $\chi(\omega,\bp)$ \eqref{eq:RPA}. The diagrams included in the RPA approximation are shown as bubbles, where $\chi^0(\omega,\bp)$ is the density-density correlator of the neutral Fermi liquid \eqref{equ:zero_sound}, and $V_\bp$ is the Coulomb interaction \eqref{eq:2}.}
\end{figure}

If the Coulomb interaction is sufficiently weak, it can be fully accounted for by  time dependent mean field
 \cite{abrikosov2012methods,nozieres2018theory}: the random phase approximation (RPA) amounting to a simple "bubble resummation" (see Fig.\,\ref{fig:RPA}).
The result is the dynamical charge susceptibility $\chi(\omega,\bp)$,
\begin{equation}
\label{eq:RPA}
\chi(\omega,\bp) \equiv \la n(p) n(-p) \ra_{V_{\bp}} {=}  \frac{\chi^0(\omega,\bp)}{1- V_{\bp} \chi^0 (\omega,\bp)} +\ldots.
\end{equation}
This has the same form as the electromagnetic charge density response with the difference that the density-density propagator of the neutral model system $\chi^0 (\omega,\bp)$ 
has replaced the polarization propagator of the charged system.

{If the neutral system has a zero sound response,
$\chi^0$ has the same structure as Eq.\,\eqref{equ:zero_sound}.} As we will see, this is also true in the case of the holographic strange metals with the difference that in that case $\Xi(\omega,\bp)$ never vanishes.
Let us therefore insert  Eq.\,\eqref{equ:zero_sound} in the expression for the  charge susceptibility \eqref{eq:RPA}. This yields 
\begin{align}
\label{eq:chi_dressed}
\hspace{-.3in}
\chi(\omega,\bp)& =
\frac{1}{1{-}V_{\bp}\bp^2\Xi} \, \frac{\bp^2A {+} \bp^2\Xi(\omega^2 {-} (v_s \bp)^2 {+} i \omega \Gamma +\frac{\Gamma^2}{4})}{\omega^2 {-} (v_s \bp)^2 {+} i \omega \Gamma +\frac{\Gamma^2}{4} {-} \frac{V_{\bp}\bp^2 A}{1{-}V_{\bp}\bp^2\Xi}}\notag \\
&= \frac{\bp^2 \tilde{A}}{\omega^2-(v_s \bp)^2  - \tilde{\omega}_p^2 + i \omega \tilde{\Gamma}}.
\end{align}
{One recognizes the plasmon excitation,} but its mass, attenuation and pole strength {receive contributions from the} incoherent second sector, 
\begin{align}
\tilde{\omega}_p^2 & 
=  
{A}V_{\bp}\bp^2 +AV_{\bp}^2\bp^4\text{Re}(\Xi)+ \ldots, \nonumber\\
\tilde{\Gamma} &
= \Gamma 
+ AV_{\bp}^2\bp^4 (-\text{Im}(\Xi)/\omega)+\ldots, 
\nonumber \\
\tilde{A} &
 = A + AV_{\bp}\bp^2\Xi +\ldots.
\label{eq:4}
\end{align}
With this parametrization, we can now isolate the signatures of the specific structure of the $\Xi (\omega, \bp)$ in the observable density response {and compare strange metals to well-known Fermi liquids}. {Inspecting Eq.\,\eqref{eq:4} we see the following.}
\begin{enumerate}[wide]
\item The Coulomb interaction promotes the $\mathbf{p} \rightarrow 0$ massless zero sound mode to the plasmon frequency $\omega^2_p$ in the longitudinal 
charge response.  In the Fermi liquid this plasmon frequency is set by the Drude weight $\omega_p^2=AV_\bp\bp^2$ alone, as $\Xi$ vanishes at $\bp=0$.\footnote{{In real metals, interband transitions can shift the plasmon frequency.}}  
In the two-sector strange metals this 
is no longer the case. There it {\em also} receives a contribution from the finite weight  $\Re \Xi(\omega,\bp=0)$ of the underlying quantum critical continuum.

\item The inverse lifetime of the plasmon $\tilde{\Gamma}$ is determined both by the "hydrodynamical" zero sound attenuation factor 
$\Gamma_{\bp} \sim \bp^2$ and by a factor associated with the spectral function $\mathrm{Im} \Xi(\omega,\bp)$.
In the Fermi liquid, this  is responsible for the Landau damping of the plasmon that only occurs at large momenta. In the two-sector strange metals this contribution 
may cause damping at all momenta. {This includes the long-wavelength limit all the way to $\bp=0$.} 

\end{enumerate} 

{We will now verify these predictions in a general class of holographic strange metals.}

\section{Computing the density response in holographic strange metals.}
\label{sec:holographic_model}

We first show how to compute the density response in holographic strange metals. 
We consider systems which at low temperature have a strongly coupled quantum critical sector determined in terms of a hyperscaling violation ($\theta$) and dynamical critical ($z$) exponents. These are readily modeled in terms of anti de Sitter (AdS) Einstein-Maxwell-Dilaton (EMD) gravity -- see \cite{Zaanen:2015oix} for a review. The $U(1)$ gauge field theory in the bulk encodes for a global $U(1)$ symmetry in the quantum critical matter system of the boundary, associated with the particle number density. Conventional holography does not take into account the (external) electromagnetic force in the boundary but we learned already in the previous section how to deal with that (see also next section).
Similar to the neutral Fermi liquid, the finite density translationally invariant holographic strange metals are invariably characterized by a zero sound mode   \cite{Karch:2009zz,Davison:2011ek,Gushterov:2018spg}
\cite{2008PhRvD..78h6004K,2009NuPhB.815..125K,2008JHEP...12..075K,2010JHEP...02..021K,2010JHEP...10..058E,2010JHEP...09..086H,2011NJPh...13g5010N,2010JHEP...11..120L,2011JHEP...12..037D,2012JHEP...10..045G,2012JHEP...10..045G,2012JHEP...11..084B,2013JHEP...06..100D,2013PhRvD..88d6010D,2013PhRvD..88h6004E,2014JHEP...02..090B,2014JHEP...04..042T,2014JHEP...04..149D,2014JHEP...07..109D,2015PhRvD..92b6004J,2016NuPhB.909..677I}. This fact was originally part of the motivation to identify these holographic models as strange metals.

\newcommand{\cA}{{\cal A}}
\newcommand{\cB}{{\cal B}}
Let us first remind the reader regarding the generalities of holographic duality.
The generating functional of the strongly coupled matter theory in the presence of a source $\cA$ is posited to be equal to the value (exponential) of the classical AdS gravitational action in one dimension higher evaluated on the solution of the equations of motion. These equations are solved with boundary condition $A(u)|_{u\rightarrow 0}=\cA$ where $u$ labels the additional dimension and $u=0$ is the location of the asymptotic boundary of the space-time. Differentiating with respect to the sources then generates the correlation functions.

Here we are interested in the density-density correlation function. As mentioned above, the number density operator $n(p)$ is dual to temporal component $A_0(p)$ of the $U(1)$ gauge field in the bulk, i.e., its boundary value $\cA_0(p)$ acts as the source for $n(p)$. In terms of the partition function of the ``boundary'' strongly coupled theory it reads
\begin{equation}
\label{equ:parti_fun}
\mathcal{Z}[\mathcal{A}] = \int \mathcal{D} \Psi \, \mathrm{exp} \left(-S_\mathrm{boundary}[\Psi] - \int \! dp \, n[\Psi] \mathcal{A}_0 \right).
\end{equation}
Applying holography, this can be equated with the (exponential of the) on-shell gravitational action 
\begin{align}
  \label{eq:3}
  \mathcal{Z}[\mathcal{A}] 
=
  \exp\left(- S_\text{AdS}[A_{\mu}(u)]\right)\Big|_{\text{on-shell};\ A_0(u=0)=\cA_0}
\end{align}
The (connected) two point function is thus obtained by taking the second variation of the on-shell action \eqref{eq:3} with respect to the sources $\cA_0$:
\begin{align}
\label{equ:variation_of_action}
\chi^0(\omega,\bp) &\equiv 
\la n(p) n(-p) \ra 
\equiv \left. \frac{\delta^2 \mathcal{Z}[\mathcal{A}]}{\delta \mathcal{A}_0(p) \delta \mathcal{A}_0(-p)} \right|_{\mathcal{A}_0 \rar 0} \notag\\
&= \left. \frac{\delta \la n(-p) \ra_{\cA_0} }{\delta \mathcal{A}_0(-p)} \right|_{\mathcal{A}_0 \rar 0}
\end{align}

Doing so one arrives at the simple result that the
two-point function can be read off from the near-boundary behavior of the solution to the equations of motion for $A_0(p;u)$. For a $U(1)$ gauge field, this is of the form
\begin{equation}
\label{equ:asympt}
A_0(p;u)\Big|_{u\rar0} = a(p) \big(1 + O(u)\big) + u^{\alpha} b(p) \big(1 + O(u)\big), 
\end{equation}
with $\alpha>0$ being the exponent of the subleading mode near the boundary, dictated by the equations of motion. 
As mentioned above, in standard holography the perturbative source just sets the Dirichlet boundary condition for the bulk field $A_0$
\begin{equation}
\label{equ:dirichlet}
a(p) \equiv A_0(p;u)\big|_{u \rar 0} = \cA_0(p).
\end{equation}
The on-shell action $S_\mathrm{AdS}$ reduces to the boundary term
\begin{equation}
\label{equ:s_onshell}
S_\mathrm{AdS}\big|_\mathrm{on-shell} = \int \! d^{D-1}p \, a(p) b(-p),
\end{equation}
therefore the expectation value $\delta S/\delta \cA$ induced by the source is associated with the subleading coefficient $b(p)$. Below we will have the case with $\alpha=1$ in Eq.\,\eqref{equ:asympt} meaning that the expectation value is given by a first derivative on the boundary
\begin{equation}
\label{eq:vev}
\delta \la n(p) \ra = b(p) = \p_u A_0(p;u)\Big|_{u\rar0}.
\end{equation}
The holographically computed two point function is therefore (see  \eqref{equ:variation_of_action})
\begin{equation}
\label{equ:b_over_a}
\chi^0(\omega,\bp) \equiv \la n(p) n(-p) \ra = \frac{b(p)}{a(p)}\,.
\end{equation}

For a unique solution one does need additional boundary conditions in the interior. 
For a black-hole space-time as we shall consider here, choosing \textit{infalling} boundary conditions at the black-hole horizon results in the retarded thermal Green's function in Eq.\,\eqref{equ:b_over_a} \cite{son2002minkowski}.

In what follows we compute the density-density correlator in the Einstein-Maxwell-Dilaton backgrounds \cite{2010JHEP...11..120L,Charmousis2010}  encoding for translationally invariant holographic strange metals with a quantum critical sector that can be tuned in a flexible manner. These have the gravitational action 
\begin{equation}
\label{equ:action}
S{=}\int \dd^{D} x \sqrt{-g} \left[\frac{R {-}2\Lambda}{2\kappa^2}{-} \frac{1}{2} (\p \phi)^2 {-} V(\phi) {-} \frac{Z(\phi)}{4q^2} F_{\mu \nu} F^{\mu \nu} \right].
\end{equation}
Without loss of generality we choose the bulk couplings for gravity $2\kappa^2=1$ and for the gauge field $q=1$. Moreover, we choose the following potentials
\begin{align}
&\Lambda=-\frac{(D-1)(D-2)}{2L^2}\,,\  Z(\phi) = e^{\gamma \phi}, \notag\\
 &\hspace{7mm}V(\phi) = \frac{2m^2}{\delta^2}\sinh^2\left({\delta \over2} \phi\right)\;.\label{equ:potentials}
\end{align}
We then solve the equations of motion numerically for $D=4$ dimensions subject to the assumption of translational invariance and \textit{without a probe source} for the density operator $\cA_0(p)$ (see Appendix \ref{app:EOMs}). We choose $m^2L^2=-2$, which fixes the  behavior of the scalar field near $u\to0$: $\phi\sim \phi_a u + \phi_b u^2+\dots$, and we impose $\phi_a=0$ as boundary conditions. All other boundary conditions may be found in \cite{Kiritsis:2015oxa}.
The resulting spacetime geometry is dual to a $2+1$ dimensional strange metal state and the presence of a quantum critical sector is exhibited through an emergent scaling IR geometry at zero temperature of the form \cite{Kiritsis:2015oxa}:
\begin{gather}\label{ds2 anstaz}
  \dd s^2 = 
{u^\theta}\left(
  - Q_{tt}^{(0)} {\dd t^2\over u^{2z}}
  + Q_{uu}^{(0)} { \dd u^2\over u^2}+ Q_{xx}^{(0)}  {dx_1^2 + dx_2^2\over u^2} \right)\,, \\
\notag
A_0 = u^{{\zeta} - z} a^{(0)}, \qquad e^\phi = u^{{\kappa}} f^{(0)},\\
\notag
  z=\frac{\gamma^2+2\delta\gamma-3\delta^2+4}{\gamma^2-\delta^2}\,,\notag \quad \theta=\frac{4\delta}{\gamma+\delta}, \\
  {\kappa} = \frac{4}{\gamma + \delta}, \quad {\zeta} = \theta-2,
\end{gather}
for some constants $Q_{tt}^{(0)}$, $Q_{uu}^{(0)}$, $Q_{xx}^{(0)}$, $a^{(0)}$ and $f^{(0)}$. Rescaling $t\to\lambda^z t\,,\ x_i\to \lambda x_i\,,\ u\to\lambda u$ the metric transforms $\dd s^2\to \lambda^\theta\dd s^2$. This shows the emergent scaling behavior with $z$ being the dynamical critical exponent of the quantum critical sector. At low temperatures the black-hole entropy density  scales as $s=T^{\frac{d-\theta}{z}}$, where $d=D-2=2$ is the number of space dimensions in the boundary theory. This identifies $\theta$ as the hyperscaling violation exponent, which is intuitively understood as parametrizing the effective space dimensionality, as shown by the scaling exponent of entropy: $(d-\theta)/z$.
In what follows we will use particular values of $z$ and $\theta$ that ensure a vanishing ground-state entropy   \cite{Kiritsis:2015oxa}. We often consider the case of large $z$ in order to make connection with previous results in the literature (numerically the largest value we use is $z=6401$). We will simply denote this value as $z\to\infty$.

\begin{figure*}[t!]
	\centering
	\hspace{-8mm}\raisebox{+.0in}{
		\includegraphics[width=0.45 \linewidth]{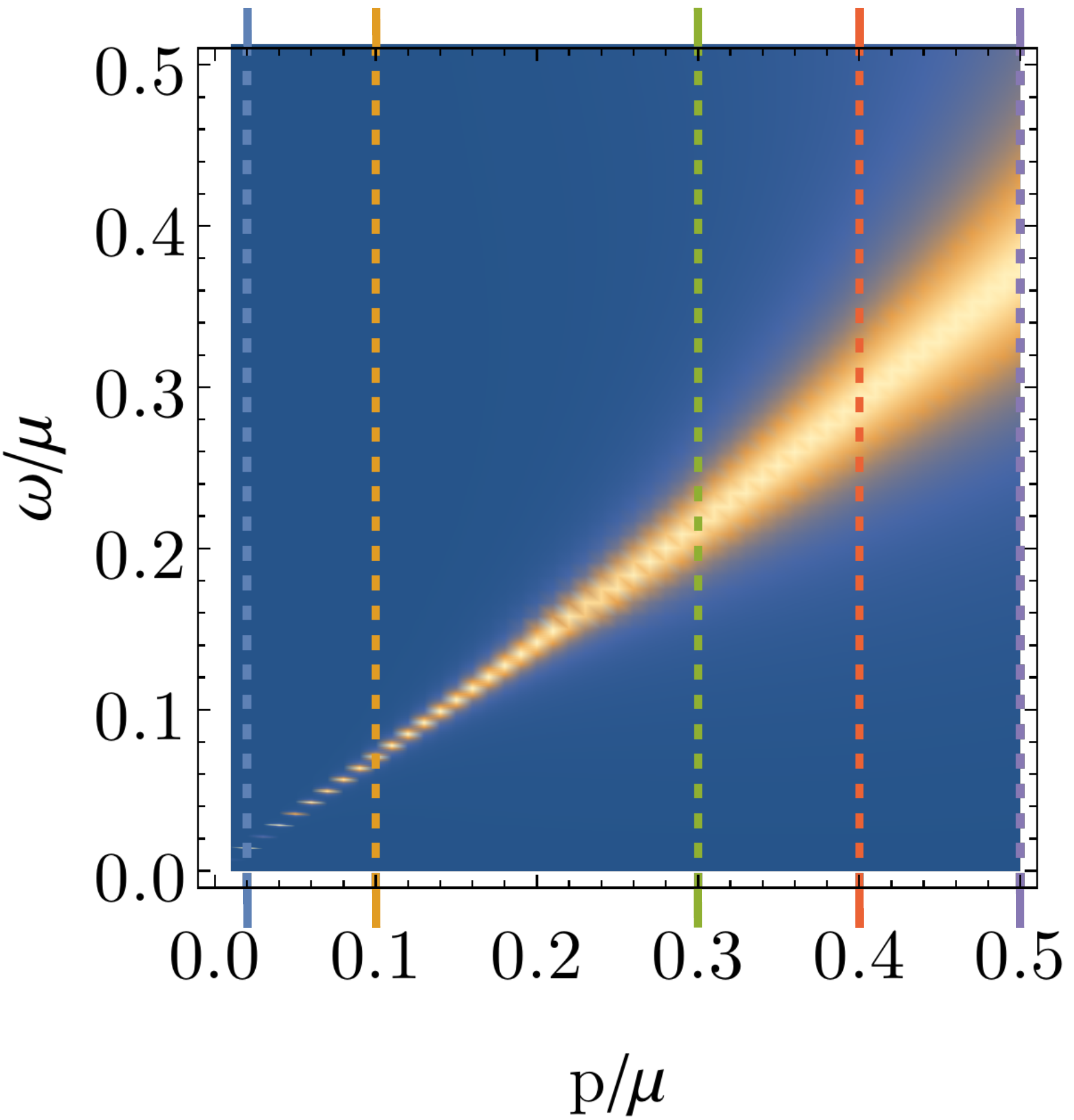}
	}
	\hspace{-3mm}
	\includegraphics[width=0.45\linewidth]{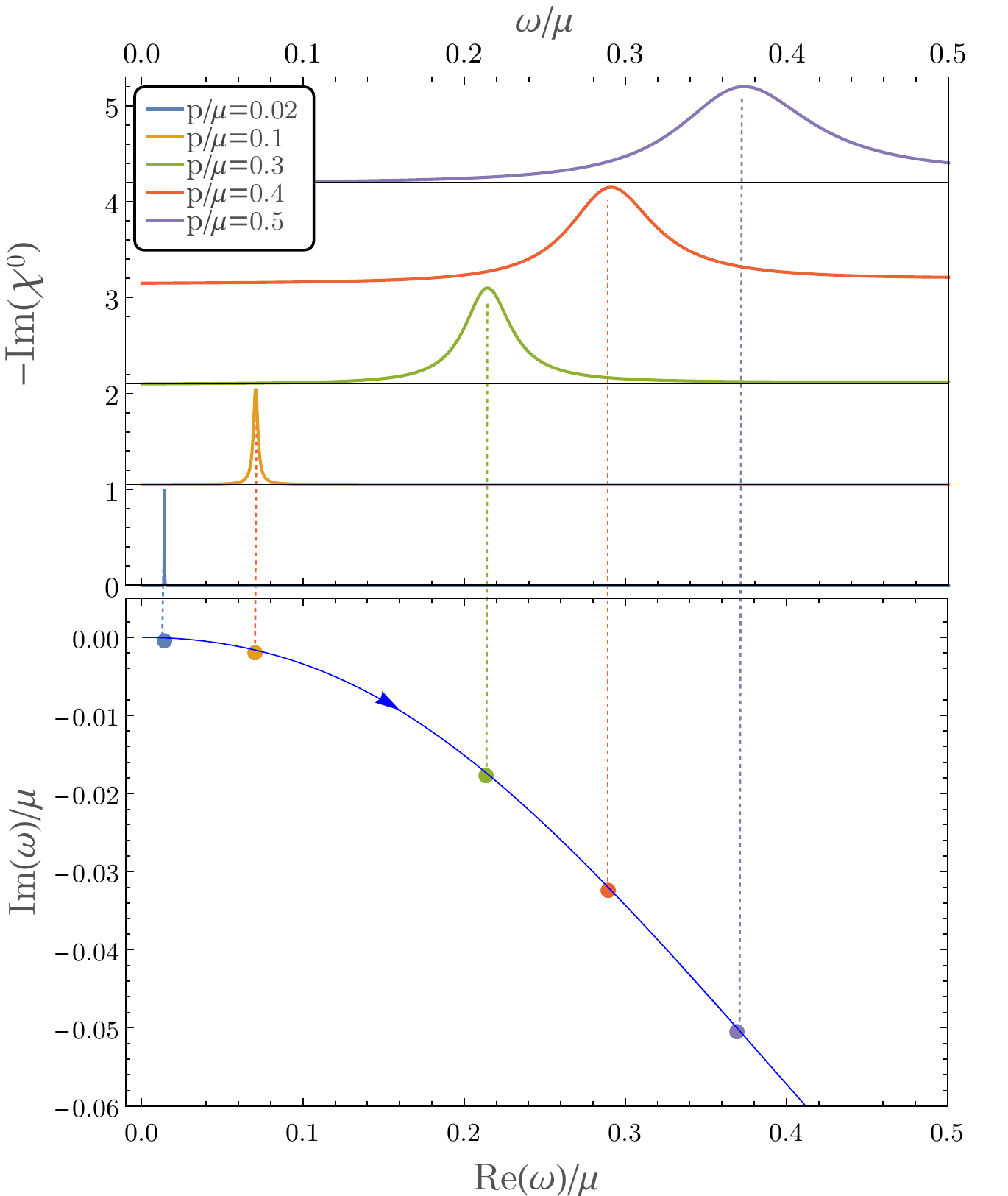}
	\caption{\label{fig:sound_QNM}
		{\bf The zero sound response in holographic strange metals.} (Left) The imaginary part of the density-density correlation function $-\Im\chi^0(\omega,\bp)$ computed in locally quantum critical strange metal ($z\to\infty,\,\theta=ent0)$ at low temperature $T\sim 0.02\mu$. The sharp zero sound response is clearly visible. (Right) The top panel shows five color-coded cuts of the spectral function (normalized to peak height). The bottom panel shows the motion of the lowest quasinormal mode in the complex $\omega$ plane as a function of momentum $\bp$. This illustrates that the zero sound response is indeed fully controlled by this single pole. }
\end{figure*}

On top of such backgrounds we then study
the linearized equations of motion of momentum-dependent perturbations induced by the source $\cA_0(p)$. This is a system of 11 coupled ordinary differential equations which we solve again numerically; details are also discussed in Appendix\,\ref{app:EOMs}. The result for the two-point function of the density operator is then extracted from the solution following Eq.\,\eqref{equ:b_over_a}. Its spectral function Im$\chi^0 (\omega,\bp)$ is plotted in Fig.\,\ref{fig:sound_QNM} for the simplest such model: a locally quantum critical strange metal with $z=\infty,~\theta=0$. One clearly sees the zero sound peak with characteristic linear dispersion and zero width in the long-wavelength limit.

There is a useful technique to analyze these coherent peaks more precisely. We study the linearized equations of motion with boundary conditions Eq.\,\eqref{equ:asympt} where $\cA_0=a=0$ (the Sturm-Liouville problem, which we further discuss in Appendix\,\ref{app:EOMs}). The frequencies that satisfy this condition specify the poles of the two-point function \eqref{equ:b_over_a} and correspond to the \textit{quasinormal modes} of the dual black hole. For a given momentum, these poles will lie in the lower half complex $\omega$ plane. The pole at (complex) $\omega_s(\bp)$ with lowest frequency is the one corresponding to the zero sound mode with the characteristics \begin{equation}
\label{equ:ReIm_QNM}
\mbox{Zero sound:} \qquad v_s \, |\bp| = \mathrm{Re} \, \omega_s(\bp) \qquad \Gamma(\bp) = -\mathrm{Im} \, \omega_s(\bp).
\end{equation}
In the bottom right plot of Fig.\,\ref{fig:sound_QNM}, we show the zero sound pole and its motion as a function of momentum $|\bp|$. We also show how this pole is reflected in the spectral function Im$\chi^0 (\omega,\bp)$ which is plotted in the top right plot of the same figure. {We clearly see that the coherent response is well described by a single pole approximation.}

At the same time, the coherent zero sound peak is not the only contribution that can be seen in the density-density correlator $\chi^0 (\omega,\bp)$. Fig.\,\ref{fig:Pi_fits} shows that there is an additional incoherent contribution
$\Xi(\omega,\bp)$ to the density-density correlation function, as we anticipated in Eq.\,\eqref{equ:zero_sound}. The quantum critical sector,  which gives this incoherent contribution, is omnipresent in holographic strange metals
and reflects the ``deep infrared scaling geometry'' near the black hole horizon in the bulk.
The effects of this second parallel sector have been studied at length in the holographic literature, especially
 in connection to {the temperature dependence of} the dc electrical  conductivity, at zero frequency and momentum where it becomes possible to compute it analytically. 
Here we observe its manifestation in the density response, which is related to the conductivity by the continuity equation [see Eq.\,\eqref{EMlinresp}]. 

{The observation that the optical response in holographic strange metals is governed by two seemingly independent systems --- one of them responsible for the Drude peak, directly associated with zero sound in our case, the other one due to the quantum critical sector --- was  made early on \cite{Hartnoll:2007ih,Blake:2013bqa,Davison:2015bea,Davison:2014lua,Davison:2015taa}. At $\bp=0$
it is easy to deduce the temperature scaling of this QC-contribution to the dc conductivity at zero momentum using general scaling arguments\cite{Karch2014,Hartnoll2015}:
\begin{equation}
\sigma_{QC} ( T, \omega{=}0, \bp{=}0) \sim T^{ d-2 + 2\Phi - \theta \over z}\,,
\end{equation}

\begin{figure}[t!]
	\centering
	\raisebox{-.3in}{
		\includegraphics[scale=0.7 ]{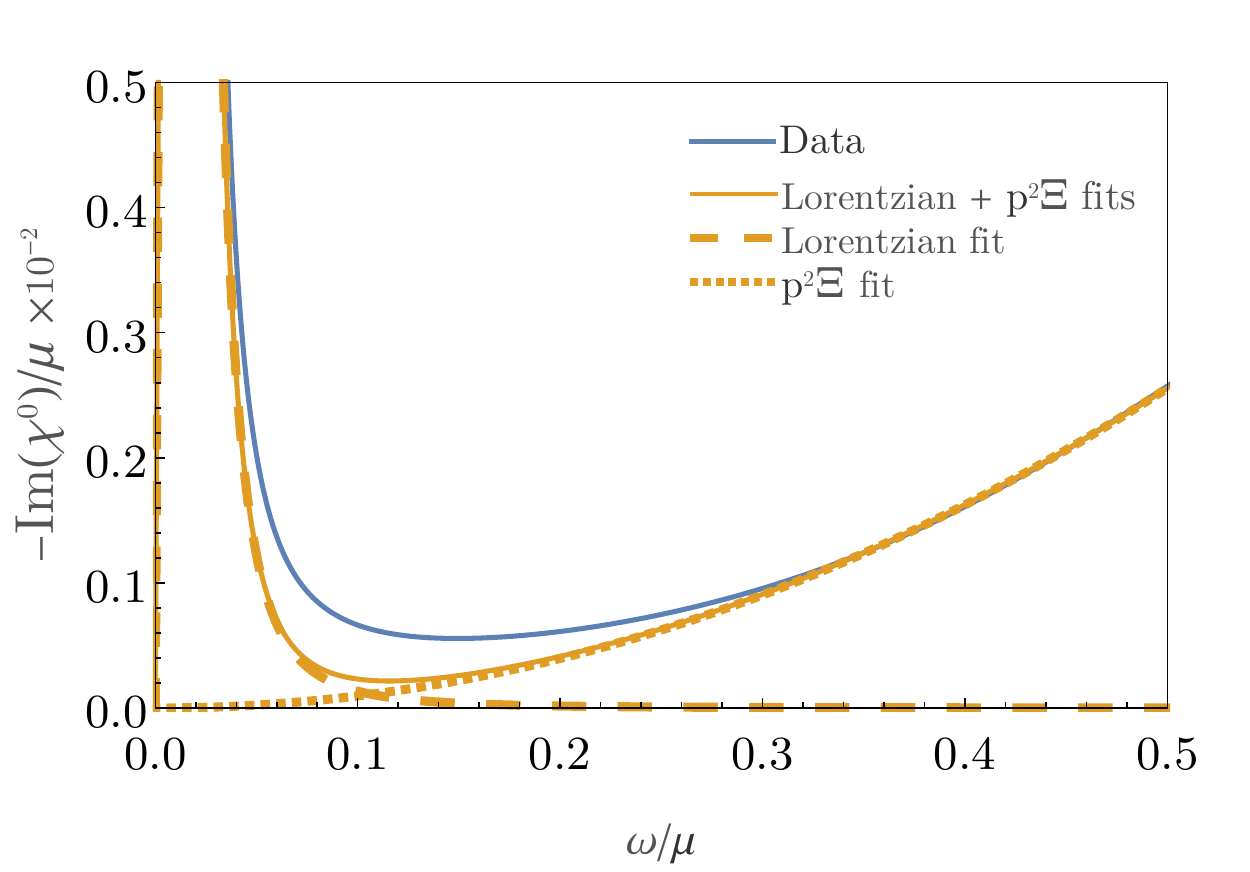} 
	}
	\vspace{-4mm}
	\caption{\label{fig:Pi_fits}
		{\bf Dissecting the density response} (continuous blue line) into its two contributions, see Eq.\,\eqref{equ:zero_sound}. The quantum critical contribution $\Xi(\omega,\bp)$ (dotted yellow) is a fit to a power law for $\omega/\mu$ between $0.3$ and $0.5$. The other contribution  (dashed yellow) is a Lorentzian fit of the sound peak. The sum of the two fits (continuous yellow) describes reasonably well the full density response. Here we take $|\bp|/\mu=T/\mu=0.02$.
	}
\end{figure}
Here $d$ equals the spatial dimensions and $z,\theta$ are the dynamical critical and hyperscaling violation exponent determined by the deep interior geometry \cite{Donos:2017mhp,Donos:2018kkm}. $\Phi$ encodes an anomalous scaling of the charge density and depends both on the geometry and on the gauge field part of the gravitational action \cite{Karch2014}. 
In our EMD model we have $\Phi=z$ always (for the absolute value, see Sec. 8.6 of \cite{Charmousis2010}); for $d=2$ this gives
\begin{equation}
\sigma_{QC} ( T, \omega{=}0, \bp{=}0) \sim T^{ \left| 3-{ \theta \over z}\right|-1}\,.
\end{equation}
The same scaling arguments then lead to the frequency scaling at zero temperature and small frequencies:
\begin{equation}
\sigma_{QC} ( T{=}0, \omega, \bp{=}0) \sim \omega^{ d-2 + 2\Phi - \theta \over z}\,,
\end{equation}
which together with Eq.\,\eqref{EMlinresp} suggests that
\begin{equation}
\Xi_{QC} ( T{=}0, \omega, \bp{=}0) \sim \omega^{ d-2 + 2\Phi - \theta -z\over z}\,.
\label{XiScaling}
\end{equation}
At finite momentum, however, scaling only fixes $\Xi$ up to an arbitrary function,
\begin{align}
  \label{eq:5}
  \Xi_{QC} (\omega,\bp) = \omega^{ d-2 + 2\Phi - \theta -z\over z} F\left(\frac{\omega}{|\bp|^z}\right).
\end{align}
In specific holographic models its asymptotic behavior near $\omega=0$ can be computed; e.g., for the class of 3+1-dimensional EMD models where $\theta \rar -\infty, z\rar \infty$ with $\eta =-\frac{\theta}{z}$ fixed, one has for infinitesimal $\omega$ \cite{Anantua2013}
\begin{align}
  \sigma_{QC}(\omega,\bp) &= i\omega\Xi(\omega,\bp) \sim  \omega^{2\nu_0}  +  \omega^{2\nu_+} +  \omega^{2\nu_-}  \ldots \nonumber\\
\nu_0 &= \frac{1+\eta}{2} \sqrt{{1+4\left(\frac{k}{V_0}\right)^2}} \nonumber\\
\nu_{\pm} &= \frac{1+\eta}{2\sqrt{2+\eta}}\left[10+\eta+4(\eta{+}2)\left(\frac{k}{V_0}\right)^2\right.\notag\\
&\hspace{16mm}\left.\pm 8\sqrt{1+(2{+}\eta)\left(\frac{k}{V_0}\right)^2}\right]^{1/2}\,,
\label{eq:7}
\end{align}
with $V_0$ proportional to the chemical potential $\mu$.

Given, however, that the plasmon frequency can be quite 
high in general (of the order of the chemical potential)} it is not clear whether these powerful deep infrared scaling forms valid near $\omega =0$ are still appropriate. Pragmatically we find that a simple
power-law form,
\begin{equation}
\label{equ:xi_ansatz}
\Xi(\omega,\bp) \sim \omega^{2\nu(\bp)}\,,
\end{equation}    
fits the numerical results for the charge susceptibility at fixed $\bp$ quite well. 

\bigskip

All along, the physical meaning of this holographic quantum critical sector has been a mystery. In the holographic community the focus has been entirely on zero 
momentum given the technical simplifications and the focus on measurable  transport properties. It is, however, quite instructive to consider finite momentum. 
It is a property of the  Fermi liquid as well. As we discussed in Section~\ref{sec:zero_sound}, 
its optical and density responses at finite momentum are characterized by a "Drude sector" (zero sound), augmented by the incoherent contributions from 
the Lindhard continuum. {As we noted below Eq.\,\eqref{3DLindhardt},}
the latter are in the deep infrared algebraic, of the same form as Eq.\,\eqref{equ:xi_ansatz}.
It is just a singular feature of the free fermion fixed point that in the Fermi liquid the $\sigma_{LC}$ disappears completely in the long-wavelength limit. However, at finite 
momentum it becomes obvious that the holographic versions should be viewed as strongly interacting (in the sense of critical theory) generalizations of the 
Fermi liquid \cite{Zaanen:2018edk}. {This is very suggestive of the way standard (bosonic zero density) CFTs can be viewed as generalizations} from the noninteracting critical state above the upper critical 
dimension. Instead of the free critical bosons found above the critical dimension, one has the richer structure of the Fermi liquid with its zero sound 
but also the algebraic (``critical") contributions associated with the Lindhard continuum. The free bosonic CFT generalizes below the upper critical dimension into the 
strongly interacting critical state characterized by universality and anomalous scaling dimensions. In a similar fashion, the Lindhard continuum generalizes in the holographic strange metal 
quantum critical infrared with its universal character.

One can push this metaphor a bit further. It is well known that in the conventional critical phenomena, universal scaling relations are violated
 above the upper critical dimension. A similar wisdom applies to the Lindhard continuum. 
{As we stressed, the frequency response of the Lindhard sector is algebraic, of the form (\ref {equ:xi_ansatz}) in space dimensions
$d > 1$. The scaling dimension $\nu$ is determined microscopically by engineering scaling: $\Xi \sim \omega$ and $\sigma \sim \omega^2$, i.e. $\nu=1$. 
On the other hand, the Fermi-liquid is characterized by $\theta = d-1$, $\Phi=1$ and $z=1$ and applying scaling analysis by extrapolating Eq.\,\eqref{XiScaling} to small but finite momentum, one would predict $\nu \sim 0$ 
instead. The actual scaling of the Lindhard continuum therefore violates the universal scaling predictions.} 

The real singular aspect of the Fermi liquid is the upper bound of the Lindhard continuum, $\omega_{max} = \frac{p_F|\bp|}{m}+\frac{\bp^2}{2m}$, responsible for its disappearance at $\bp =0$. This is intimately linked to 
the free quasiparticle  kinematics of the FL fixed point. Dealing with the ``unparticle'' physics associated with the strongly interacting 
holographic strange metals such an upper bound is completely unnatural. 
Recognizing {this unique particle} feature of the free fermion fixed point, it is obvious 
that in anything else (the ``non-Fermi liquid'') the continuum is expected to survive at zero momentum and finite 
frequency. This has the the ramification implied by the universal linear response theory of Section~\ref{sec:zero_sound} that the long-wavelength plasmon will 
have a finite lifetime. 
{This we will now show by explicit computation in holographic strange metals as a proof of principle in expectation that two-sector
quantum critical strange metals of this type are realized in nature.}

\section{ Switching on the Coulomb interaction: plasmons in strange metals.}
\label{sec:plasmon} 

We first address how to incorporate the effects of the long-range Coulomb interaction $V_{\bp}$ in the nominally neutral 
holographic strange metal.  We do so in exactly the same way as reviewed in Sec.\,\ref{sec:zero_sound}, by coupling the density as part of a $U(1)$ current to the electromagnetic field in the boundary, and integrating out the Maxwell potential followed by the nonrelativistic limit to obtain the partition function [see \eqref{eq:2}], 
\begin{align}
\label{equ:parti_fun_deformed}
\mathcal{Z}[\mathcal{A}]_{V_{\bp}} &= \int \mathcal{D} \Psi \, \mathrm{exp} \left( - S_\mathrm{boundary}[\Psi] - \int \! dp \, n[\Psi] \mathcal{A}_0 \right. \notag\\
&\left.+  \int \! dp  \, \frac{1}{2} V_\bp n[\Psi]   n[\Psi] \right).
\end{align}

This is known in the context of holography as a ``double-trace deformation'' --- it is quadratic in the ``single trace'' operator $n[\Psi]$ \cite{witten2001multi}.\footnote{A similar method of incorporating the Coulomb interaction is used in \cite{Mauri:2018pzq}.}
The holographic counterpart, the on-shell action, must be deformed correspondingly. Keeping in mind that the subleading coefficient $b(p)$ in \eqref{equ:asympt} is responsible for the expectation value of $\langle n(p)\rangle$, one adds the double-trace boundary term \cite{muck2002improved} in terms of $b(p)$,
\begin{equation}
\label{equ:S_V}
S_{V_{\bp}} = S_{AdS} - \int \! d^{D-1}p \ \frac{1}{2} \, V_\bp b(p) \, b(-p) \,.
\end{equation}
This extra term modifies the boundary conditions for the gauge field $A_0(p;u)$, which for $\alpha=1$ take the mixed form \cite{witten2001multi} (Note that for $V_\bp =0$ it reduces to \eqref{equ:dirichlet})
\begin{equation}
\label{eq:Robin}
a(p) - e^2 V_\bp b(p) \equiv A_0(p;u)\Big|_{u\rar 0}\hspace{-3mm} - e^2 V_\bp \p_u A_0(p;u)\Big|_{u\rar 0}  \hspace{-3mm}= \mathcal{A}_0(p).
\end{equation}
The perturbative source in this case is not directly related to the leading coefficient $a(p)$ as in Eq.\eqref{equ:dirichlet} and therefore the variation with respect to $\cA_0$ \eqref{equ:variation_of_action} brings down extra contributions from 
double-trace boundary term \eqref{equ:S_V}. Keeping in mind that for a given horizon boundary condition $b(p)/a(p) = \chi^0(\omega, \bp)$ \eqref{equ:b_over_a} one arrives at the result for the ``dressed'' charge density two-point function \cite{witten2001multi,muck2002improved,Zaanen:2015oix}
\begin{align}
\chi(\omega, \bp) &\equiv \la n(p) n(-p) \ra_{V_\bp} = \left. \frac{\delta^2 S_{V_\bp}}{\delta \cA_0 \delta \cA_0} \right|_{\cA_0 \rar 0} \notag\\
&= \frac{\chi^0 (\omega,\bp)}{1 -  V_{\bp} \chi^0 (\omega,\bp)},\label{equ:chi_holography}
\end{align}
which coincides exactly with the RPA formula Eq.\eqref{eq:RPA}. In Appendix\,\ref{sec:Ax} we discuss the generalization of this expression to the fully relativistic case.

\subsection{The quantum critical sector and the plasmon.}

\begin{figure}[t!]
	\centering
	\hspace*{-.2in}
	\includegraphics[width=0.545\linewidth]{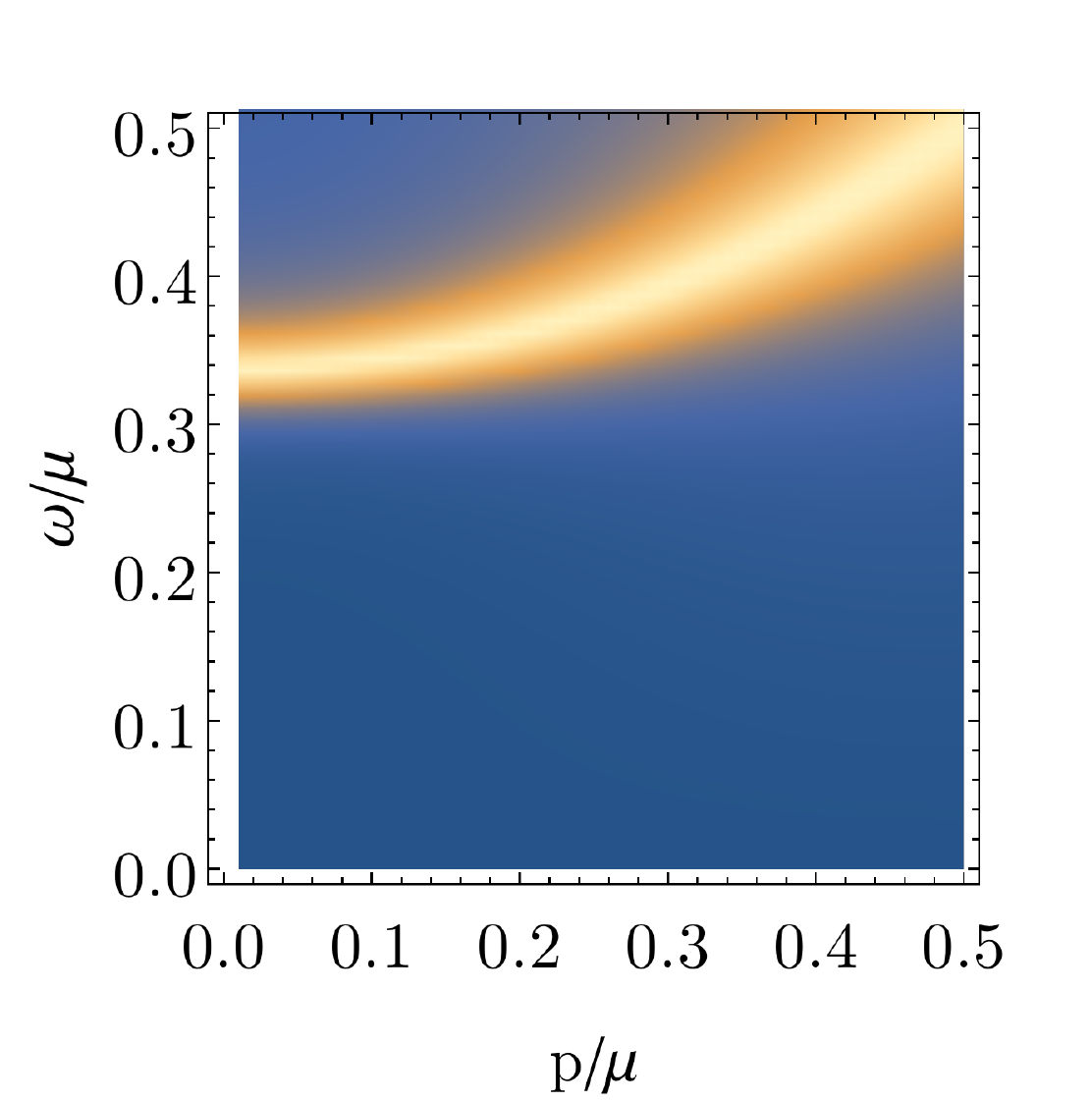}
	\hspace*{-.25in}
	\includegraphics[width=0.545\linewidth]{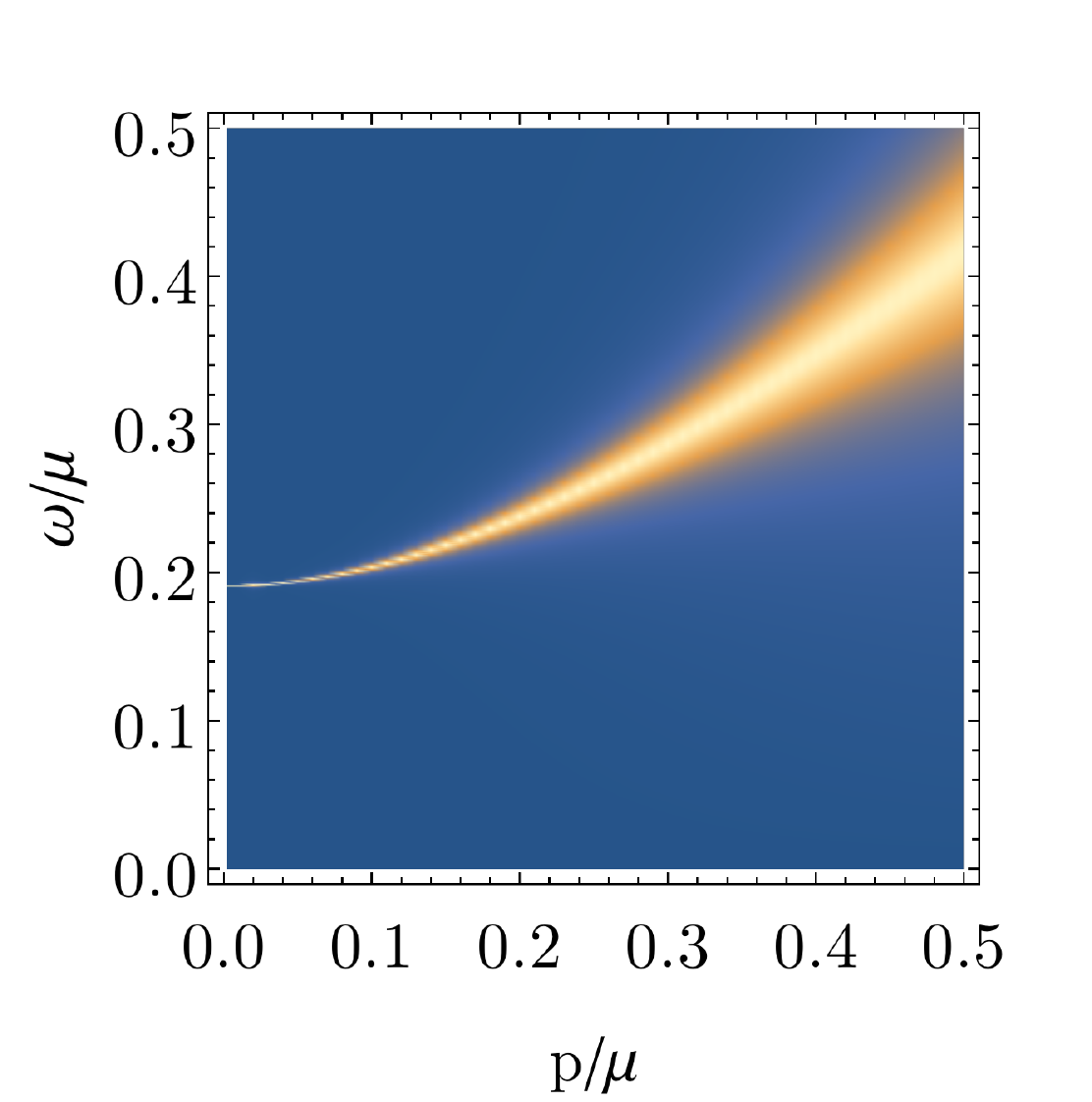}
	\raisebox{.34in}{
		\hspace*{-.3in}
		\includegraphics[width=0.06\linewidth,scale=1]{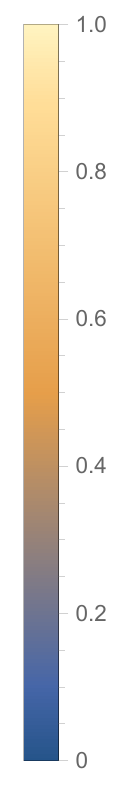}}
	\caption{\textbf{The plasmon response in holographic strange metals: the actual spectral function (Left) and the one with the quantum critical sector artificially removed (Right)}. The left plot shows the  spectral function $-\Im\chi(\omega,\bp)$ of the full dressed response $\chi(\omega,\bp)$ computed according to Eq.\,\eqref{equ:chi_holography} for the locally quantum critical strange metal ($z\to\infty,\theta=0$) and Coulomb potential $V_{\bp}=2.5/\bp^2$. In the right plot, the quantum critical sector in the same model is removed from the neutral density response by hand by isolating the pole; namely we replace $\chi^0(\omega,\bp)$ in Eq.\,\eqref{equ:chi_holography} by the  ``pole-only'' contribution shown in Eq.\,\eqref{eq:pole-only}. 
		Comparing both plots clearly shows how the quantum critical continuum leaves its imprint in the plasmon both by broadening at zero momentum and by a shift of the plasmon gap. The density plots have been normalized by the peak value at every momentum and we take $T= 0.02 \mu$. }\label{fig:dispersion}
\end{figure}

To elucidate the gross features of plasmons in holographic strange metals let us consider for the first example the most familiar case in the literature, the $z \rar \infty$ locally quantum critical metal, dual to the simple charged black hole of the Einstein-Maxwell theory in the bulk. 
Although the strict limit is pathological as it has a finite ground-state entropy (see Sec. \ref{sec:holographic_model}),
this has been studied in particular because of its comparatively simple bulk structure. 
With the number density correlator $\chi^0(\omega,\bp)$ computed numerically in the previous section in hand, we evaluate Eq.\,\eqref{equ:chi_holography} and 
the result for the imaginary part of $\chi(\omega,\bp)$ is shown in  Fig.\,\ref{fig:dispersion}.  We have chosen a value of the electromagnetic coupling $e^2 = 2.5$ in the Coulomb potential $V_{\bp}=e^2/\bp^2$ such that the plasmon frequency is a reasonable fraction of the chemical potential, of the kind that is typically encountered
in condensed matter systems (e.g., in the cuprates $\omega_p \simeq 1$ eV for a bare Fermi energy $\simeq 2$ eV). As explained, we indeed recover that the sound pole of 
the neutral system (Fig.\,\ref{fig:sound_QNM}) gets promoted to a finite energy (plasmon frequency) in the long-wavelength limit. It should now not be surprising that in contrast to the sound pole
the \textit{plasmon is damped substantially already in the long-wavelength limit}: this is the effect of the ``quantum critical damping'' of the plasmon, decaying into the second
quantum critical infrared sector.

To highlight the role of the quantum critical infrared sector,  we can artificially exclude its continuum contribution from the holographic neutral density correlator $\chi^0(\omega,\bp)$. As we discussed in the previous section, the pole structure of $\chi^0(\omega,\bp)$ can be obtained 
from the study of the quasinormal modes 
shown in Fig.\,\ref{fig:sound_QNM}, right panel.
The dominant contribution comes from the pole corresponding to zero sound $\omega_s(p)$,
and we can use this to construct a  ``zero sound pole-only'' density response with no continuum contribution $\Xi_{\text{pole-only}}(\omega, \bp) =0$:
\begin{equation}
\label{eq:pole-only}
\chi^0_{\text{pole-only}}(\omega,p) \sim \frac{1}{\omega^2 - \big[\ |\omega_s(p)|^2 {+} i 2\omega \; \mathrm{Im} \, \omega_s(p)\ \big]}. 
\end{equation}
Substituting the pole-only density response into the RPA formula \eqref{equ:chi_holography}, and computing the spectral weight of the corresponding charge density correlator, one gets the result shown on Fig.\,\eqref{fig:dispersion}, right panel. One immediately sees both our predictions  verified. Removing the quantum critical continuum (a) shifts the gap at zero momentum (similar to 
``undoing'' the effect of interband transitions in conventional systems),
and (b) reduces the decay width of the plasmon dramatically and eventually kills it completely at zero momentum. At $\bp\rar0$  only the quantum critical continuum $\Xi_{QC}(\omega,\bp)$ contributes to the width of the plasmon. 

\begin{figure}[t]
\hspace{-4mm}
    \centering{\includegraphics[scale=0.45]{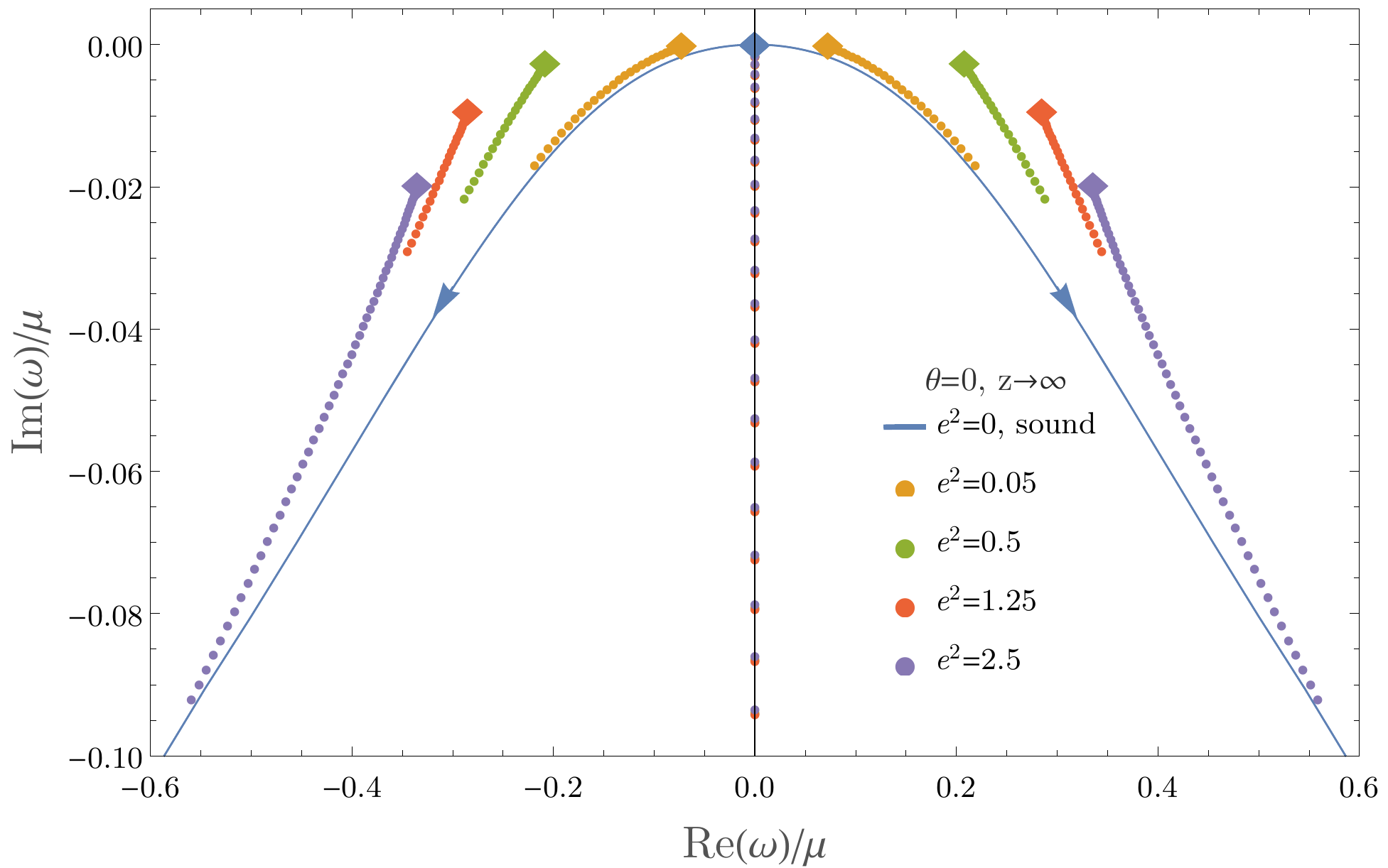}}
   \caption{{\bf Plasmonic quasinormal modes for various electromagnetic dressing factors $V(\bp)=e^2/|\bp|^2$ in the locally quantum critical strange metal ($\theta=0,\ z\to\infty$ with $T/\mu\simeq0.02$).} The solid line shows the position of zero sound modes in absence of dressing. These QNMs have been obtained with mixed boundary conditions Eq.\,\eqref{eq:Robin}. The diamonds indicate the position of the QNM at $\bp= 0$; arrows indicate movement of the QNM as momentum increases. }\label{fig:qnms_var_lambda}
\end{figure}

\begin{figure*}[t]
\hspace{-4mm}(A)
\hspace{-2mm}
\includegraphics[width=0.45\textwidth,scale=0.65]{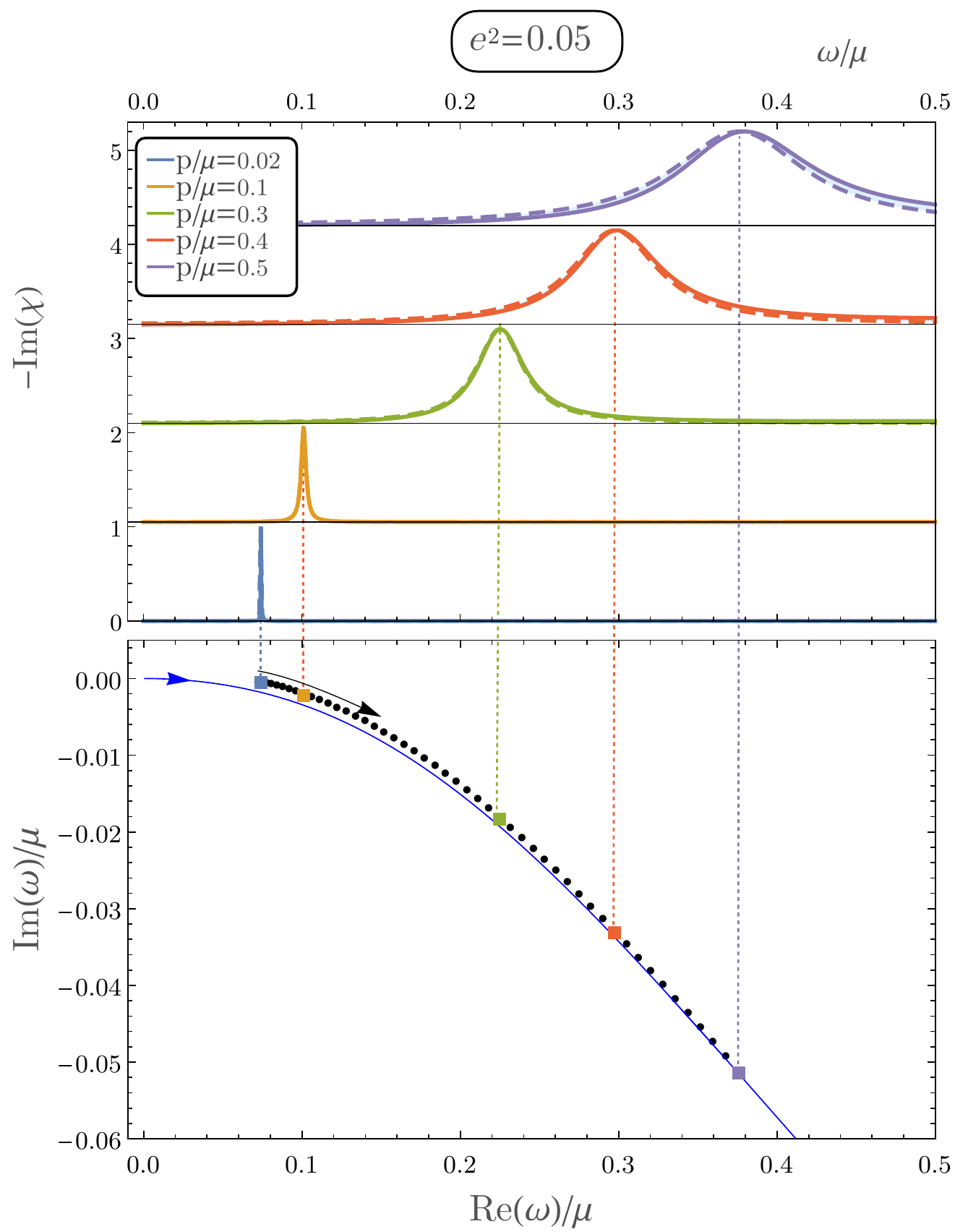}
~(B)
\hspace{-1mm}\includegraphics[width=0.45\textwidth,scale=0.65]{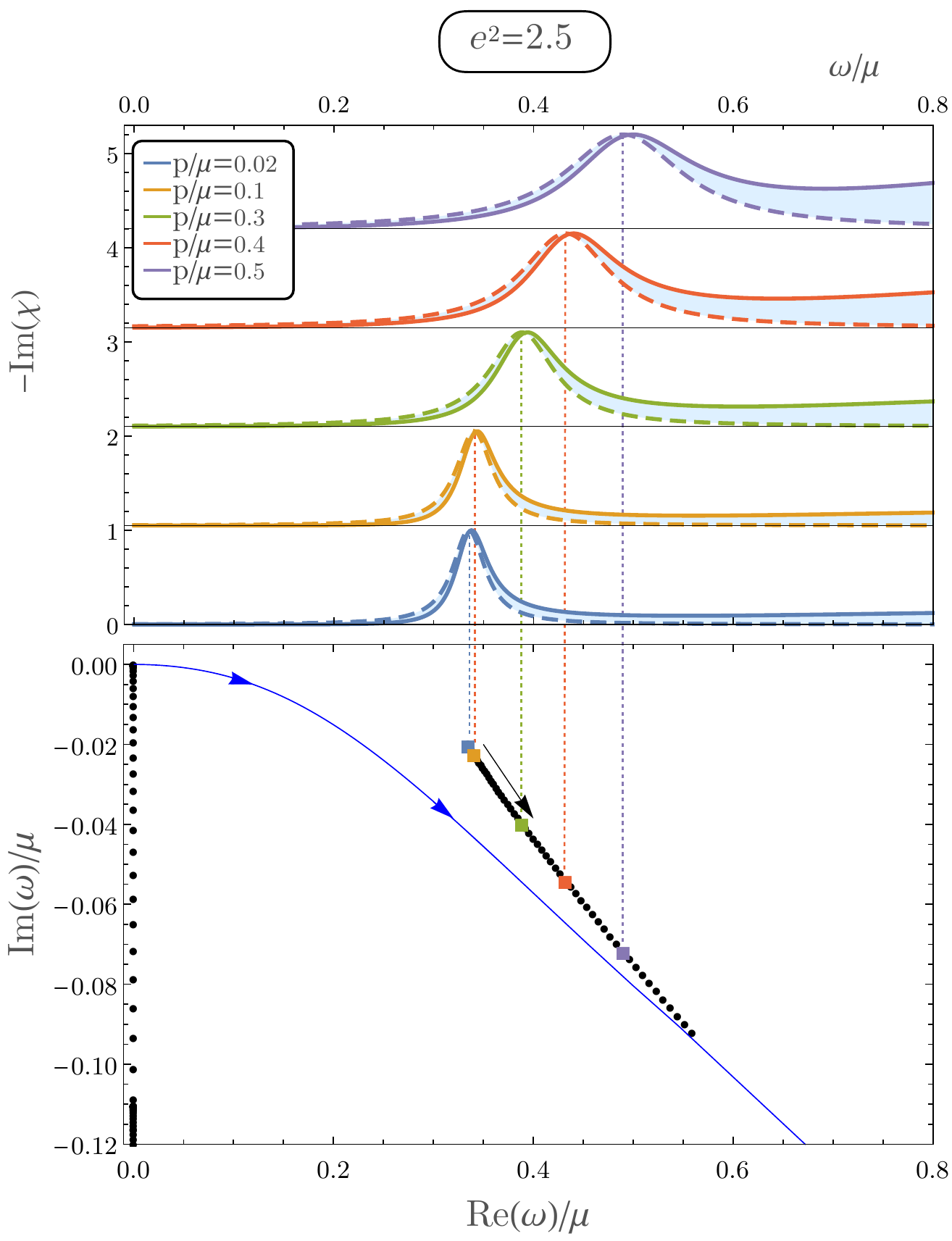}
\caption{{\bf The charge density response for fixed momentum for small (A) and large (B) Coulomb potential.} 
In the top figures the continuous lines are the fully dressed correlator $\mathrm{Im}\chi(\omega, \bp), $ Eq.\,\eqref{equ:chi_holography}, of the locally quantum critical strange metal ($T=0.02\mu, \, z\to\infty, \, \theta=0)$ as a function of frequency at different momenta. These should be contrasted with the dashed lines corresponding to the Lorentzian peak obtained from the plasmon QNM  \eqref{eq:chi_pole-only} alone. The actual spectral function is clearly non-Lorentzian and this is a signature of the presence of the quantum critical sector.
The bottom figures show the trajectories of this plasmon QNM for increasing momentum: blue lines correspond to the sound mode obtained from Dirichlet boundary conditions \eqref{equ:dirichlet} and the black dots correspond to the plasmon QNM with mixed boundary conditions \eqref{eq:Robin}.
The colored squares refer to the momentum indicated in the labels. 
The dashed  vertical lines are a guide to the eye indicating the position of the Lorentzian peaks. 
We have normalized the data by the maximum of peak and offset the different momentum slices for readability.}\label{fig:EDC}
\end{figure*}

Similar to the case of zero sound, discussed above, one can understand the plasmon response $\chi(\omega,\bp)$ in terms of the pole structure of the ``RPA-dressed'' correlator in the complex-$\omega$ plane. 
These ``plasmon QNM modes" are different from the formal quasinormal modes of the black hole as one uses different boundary conditions \eqref{eq:Robin}. Nevertheless, they are closely related and we use similar techniques to determine them; see Appendix \ref{app:EOMs} for more details. In Fig.\,\ref{fig:qnms_var_lambda} we show the momentum dependence of several sets of plasmon QNM modes, obtained for different values of the electric charge $e^2$. Again, it is immediately seen that, unlike the sound poles, the plasmon poles do not approach the origin as momentum tends to zero. Instead they halt at points with finite real and imaginary parts, which correspond to the plasmon gap and width at zero momentum, respectively. The absolute values of both real and imaginary parts grow with $e^2$, which is again expected from the RPA result \eqref{eq:chi_dressed}.

There is one more way that the quantum critical sector reflects itself in the plasmon response. Given the presence of other degrees of freedom than sound at finite
frequency, the single pole approximation with its hydrodynamical damping $\sim \Gamma_\bp$ falls short of capturing the physics well. This reflects itself also in the line shape of the 
plasmon: it is no longer a Lorentzian indicative of a long-lived mode with a finite lifetime [see  Eq.\,\eqref{eq:chi_dressed}].  The effect of the finite imaginary part 
of  $\Xi(\omega,\bp = 0)$ on the line shape is illustrated in Fig.\,\ref{fig:EDC}. 
The dashed lines show the Lorentzian peaks corresponding to the position of the leading plasmon mode $\omega_p(p)$ in the complex-$\omega$ plane

\begin{equation}
\label{eq:chi_pole-only}
\chi_{\substack{ \text{ plasmon-\ }\\ { \text{pole-only}}}}(\omega,p) \sim 
\frac{1}{\omega^2 {-} [\ |\omega_p(p)|^2 {+} i2 \omega \mathrm{Im} \, \omega_p(p)\ ]}, 
\end{equation}
whereas the actual response is shown in the continuous line. For large electromagnetic coupling $e^2=2.5$ the effect becomes especially pronounced. In particular the peak is now clearly asymmetric and has a long heavy tail towards the UV.

\section{ Plasmons and the family of holographic strange metals.}
\label{sec:tuned}

As we discussed in Section \ref{sec:holographic_model}, by tuning the potentials in the bulk EMD theory we can study a large variety of holographic strange metals, characterized by vastly
different scaling behaviors in terms of the  dynamical critical exponent $z$ and the hyperscaling violating parameter $\theta$ {}\eqref{XiScaling}. 
In this section we explore this wide parameter space and study the observable effects in the charge density response as the features of the quantum critical sector change. In principle we can compute the density response in holographic strange metals for any finite values of ($z$, $\theta$). Two subfamilies are of particular interest, however. In the family with $\theta=0$ one can study the clean dependence on the dynamical critical exponent $z$. Despite its limited applicability, we can already infer from the scaling analysis Eq.\,\eqref{XiScaling} that decreasing $z$ will decrease the contribution from the QC sector (recall that $\omega \ll \mu$.). Studying the dependence on $z$ is therefore a more fine-grained test of the QC contribution to the density response. The other subfamily of interest is the set where $z \rar \infty, \, \theta \rar -\infty$ with $\theta/z = -\eta$ fixed. This subfamily describes locally quantum critical theories ($z\rar \infty$) with tunable scaling of the entropy $s \sim T^{(d-\theta)/z}\sim T^{\frac{d}{z}+\eta}$ and includes in particular the phenomenologically interesting strange metallic scalings with Sommerfeld entropy $s \sim T$ and linear in temperature resistivity $\rho \sim T$ \cite{Hartnoll2015,Anantua2013}.

\begin{figure*}[t]
\vspace{-5mm}
   \hspace{-2mm}{\includegraphics[scale=0.8]{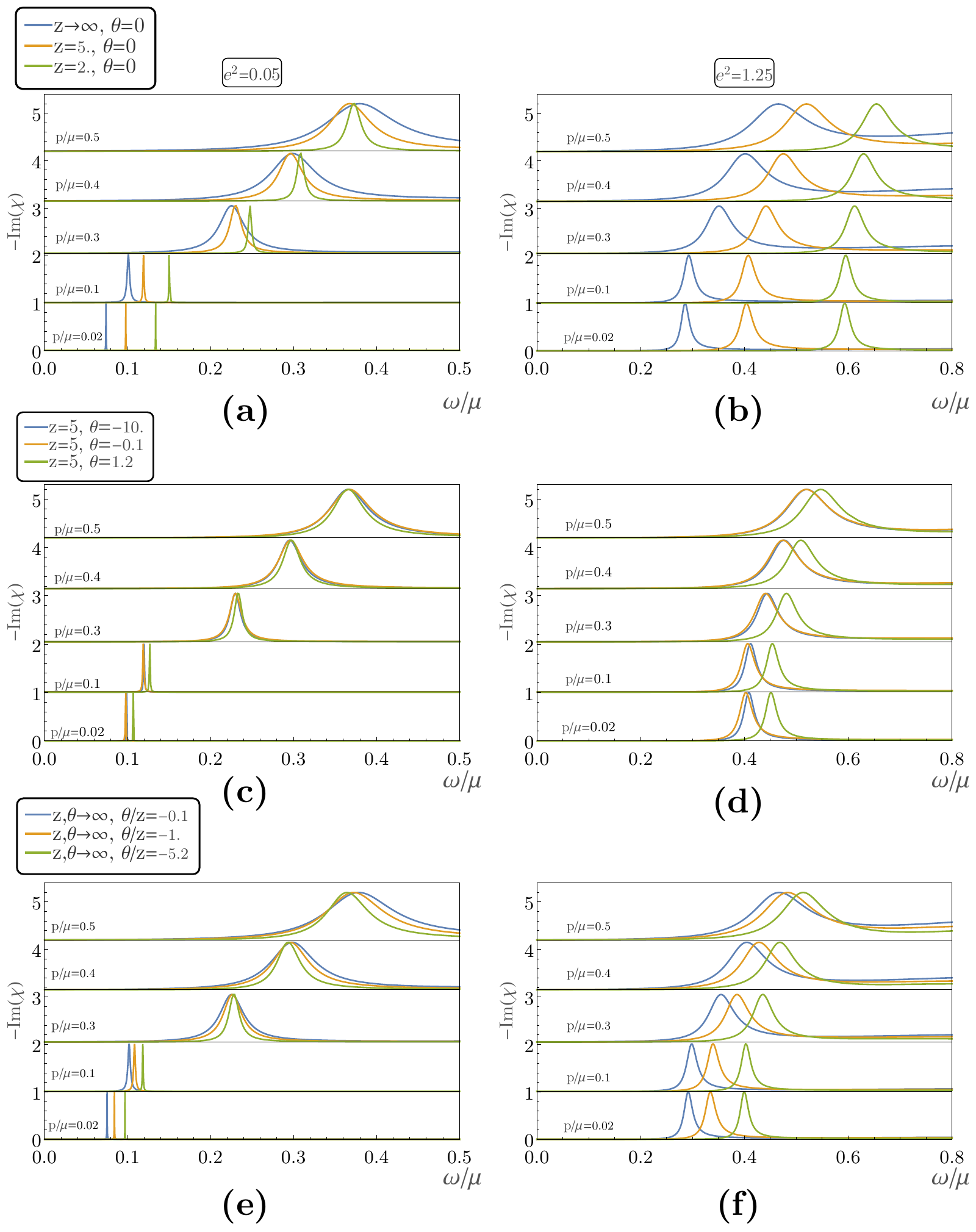}}
   \vspace{-5mm}
   \caption{Momentum slices for the dressed charge density correlator $-\Im\chi(\omega,\bp)$ at weak (left column) and intermediate (right column) coupling of Coulomb interaction and various dynamical critical $z$, and hyperscaling violating $\theta$ exponents and at  $T/\mu\simeq0.02$. We show families with $\theta=0$ (first row), generic finite $z$ and $\theta$ (second row) and a special case with $z,\theta \rar \infty$, $\theta/z$ finite (third row). For weak Coulomb coupling, as seen in Eq.\,\eqref{eq:4}, the plasmon width is dominated by the hydrodynamic sound attenuation $\Gamma \sim \bp^2$. For stronger  coupling, the contribution from quantum critical continuum $\Xi(\omega,\bp)$ is important and it determines the plasmon width completely at zero momentum. Surprisingly for all the backgrounds these widths at small momentum look identical.}
   \label{fig:z_dependence}
\end{figure*}

In Fig.\,\ref{fig:z_dependence} we show our results for the charge density response in these special families (the first and the third rows) and the generic case (second row). The left column corresponds to weak electromagnetic dressing $e^2 = 0.05$, while the right one shows the results with $e^2 = 1.25$ of order unity.

Let us focus on the weak dressing case first (left column of Fig.\,\ref{fig:z_dependence}). It follows from Eq.\,\eqref{eq:4} that in this case the contribution of quantum critical continuum $\Xi$ is small and the plasmon decay is essentially completely governed by the hydrodynamical-like damping $\Gamma ( \bp)$. Indeed we observe in Fig.\,\ref{fig:z_dependence} that for all these families the width of the plasmon is almost zero in the long-wavelength limit and grows with momentum. 
We do see, however, that the value of $\Gamma(\bp)$ does depend on the scaling features of our model. It can be understood in the following way. 
{
At low temperatures decreasing $z$ from $z=\infty$ or increasing $\theta$  suppresses the contribution from the quantum critical sector not only to  $\Xi(\omega,\bp)$ but also to $\Gamma(\bp)$. In holographic models this is computationally clear as the diffusivity is controlled by a black-hole horizon in the scaling geometry. 
The computed physical response at finite $\bp$ reflects this as} the plasmon width clearly increases as $z$ grows (Fig.\,\ref{fig:z_dependence}-(a)). 
On the other hand in the family with $z,\theta\to\infty$ (Fig. \,\ref{fig:z_dependence}-(c)) the plasmon decay rate increases with reducing the absolute value of $\eta = -\theta/z$.  
In essence for larger $z$ and smaller $\theta$,
the plasmon has more channels into which it can decay and its width grows.

\begin{figure*}[t]
	\hspace{-2mm}{\includegraphics[scale=1]{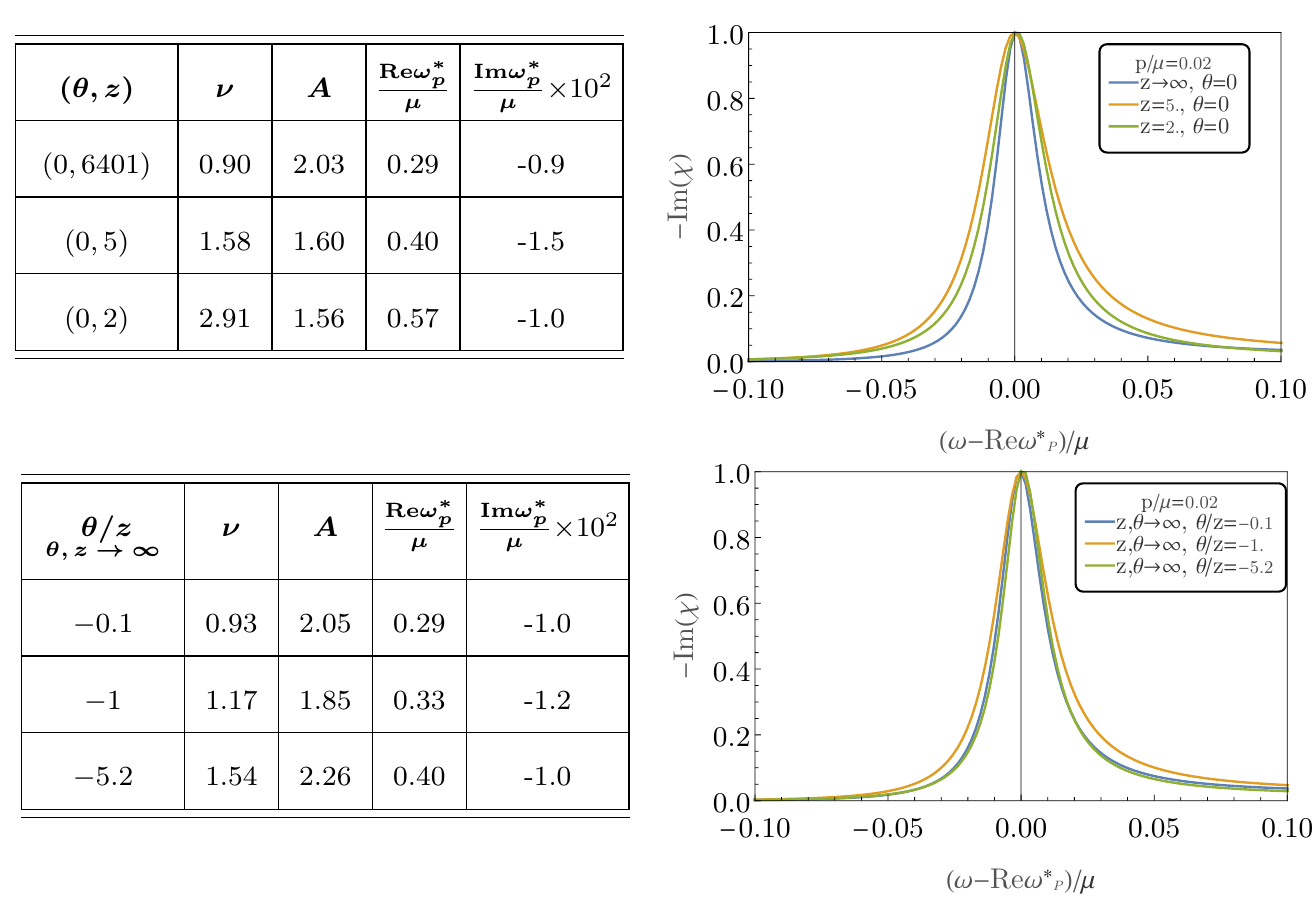}}
	\vspace{-4mm}
	\caption{The table shows the relevant parameters resulting from fitting $\chi^0(\omega,\bp)$ to Eq.\,\eqref{equ:zero_sound}  with $\Xi=a\omega^\nu$. $\omega^*_P$ is the pole of the analytical expression for $-\Im(\chi)$, which results from dressing the fitted approximation to $\chi^0(\omega,\bp)$ using  Eq.\,\eqref{eq:RPA}.
		The figures on the right show the normalized dressed response from Fig.\,\ref{fig:z_dependence}-(b),(d) at $|\bp|=0.02\mu$, shifted by the position of the maximum. The plasmon widths are remarkably similar and their dependence on the anomalous exponents follows the same trend as $\Im\omega^*_P$ shown in the tables. It becomes clear that the plasmon width is affected by both the features of the quantum critical sector and the sound pole residue: $A$. These contributions  compensate each other such that little dependence on the details of the sector remains.}\label{table}
\end{figure*}

Next we examine a larger coupling $e^2=1.25$( see right column of Fig. \ref{fig:z_dependence}). Here the dressed response acquires a sizable contribution from the quantum critical continuum $\Xi(\omega,\bp)$ [see Eq.\,\eqref{eq:4}]. This is particularly pronounced at zero momentum, where $\tilde{\Gamma} \rar 0$ and the width of the plasmon is completely dominated by $\Xi(\omega,\bp)$. In this long-wavelength case we can study the continuum contribution very cleanly.
As discussed above in Sec.\,\ref{sec:holographic_model} we expect $\Xi(\omega,\bp)$ to be strongly dependent on the $z$ and $\theta$ exponents \eqref{XiScaling}. 
However, rather surprisingly, we observe that the widths of the peaks at small momenta are not very sensitive to our tuning parameters. 
There are only small differences which can be seen by superposing the peaks (see Fig.\,\ref{table}). 

From a more careful examination we understand why this sensitivity is not present in the response. First we note that the expression for the width of the plasmon at zero momentum, $\tilde{\Gamma}(0) \sim - A \text{Im}(\Xi)/\omega $, receives contributions from {\em both} the value of the QC continuum at plasmon frequency $\omega_p$ and the residue of the sound pole $A$. In order to figure out the interplay between these contributions we extract these parameters from a fit to the neutral ``undressed'' density susceptibility $\chi^0(\omega,\bp)$ in the different models under consideration. We use the same ansatz \eqref{equ:zero_sound} and obtain the fits as shown in Fig.\,\ref{fig:Pi_fits}. The values we obtain for $A$ and the QC power law exponent $\nu$ at small momentum $\bp=0.02\mu$  [see ansatz \eqref{equ:xi_ansatz}] are shown in Fig.\,\ref{table}. 
When we substitute these fit-obtained values in the expression for the dressed correlator $\chi(\omega,\bp)$ \eqref{eq:chi_dressed} we can extract the position and the width of the plasmon pole by finding the zero of the denominator numerically. These we show as Re$(\omega^*_p)$ and Im$(\omega^*_p)$ in the table of Fig.\,\ref{table}.
We see that these values describe rather well the observed width of the peaks in the directly computed full correlator $\chi(\omega,\bp)$, shown in the right panel of Fig.\,\ref{table}, but also that they barely change as we change $z$ or $\theta$. This confirms that our ansatz correctly takes into account this low sensitivity issue.

Inspecting more closely one sees that it is true that, as expected from Eq.\eqref{XiScaling}, the continuum power law exponent $\nu$ depends strongly on the background scaling features.\footnote{Note, however, that the exact value of the fitted power law which we find does not coincide with the prediction of Eq.\eqref{XiScaling}. The reason is that the former is fitted in the range of frequencies of order $\omega_p$ (see Fig.\,\ref{fig:Pi_fits}), while the latter is only valid in the limit $\omega \rar 0$.}  
However, the sound pole residue $A$ also changes. It contributes {\em oppositely} in such a way that it
 stabilizes the plasmon width. This is surprising, though in a qualitative sense one should know that $A$ and $\nu$ are not entirely independent. There is a nontrivial connection between $A$ and $\nu$ arising from the spectral sum rule for the number density correlator (where we use the ansatz \eqref{eq:chi_dressed})
 \begin{align}
\int\limits_{-\infty}^\infty \hspace{-2mm}
\dd \omega\  \omega \Im \chi^0 &{=}{-}|\bp|^2 A\pi {-}|\bp|^2\hspace{-1mm} 
\int\limits_{-\infty}^\infty \hspace{-2mm}\dd \omega\    \omega \Im\Xi = {\rm const}\times |\bp|^2.\label{eq:sum_rule}
\end{align}
Even though Eq.\,\eqref{eq:sum_rule} does not prescribe the exact relation between $A$ and $\nu$, it correlates the changes in $\Xi(\omega,\bp) \sim \omega^\nu$ to changes of  $A$. {Nonetheless, that they counteract each other so closely is surprising.}

\section{Conclusions}

As is often the case, holography has a remarkable capacity to suggest unusual general physics questions of special relevance to condensed matter \cite{Zaanen:2015oix} 
that in hindsight have simple answers. The present story is an illuminating illustration. 

The impeccable, long-lived nature of the plasmon ``particle'' is at the heart of the large engineering pursuit of plasmonics. It should be clear by now that this 
is actually rooted in the special, singular {free-particle} nature of the Fermi-liquid fixed point. The existence of sound is a different matter -- it is a consequence 
of conservation laws and although there is no mathematical proof that it has to exist at zero temperature it may well be ubiquitous. By the same token, given 
the universal nature of the electromagnetic linear response theory, this massless sound will be promoted to the plasma frequency at zero momentum. 
However, for the plasmon to have an infinite lifetime at this finite energy, it has to be completely isolated from all other energy eigenstates that couple 
to charge density. There is no general, symmetry based argument insisting that this is a natural condition. The very special nature of the 
Fermi liquid plays the primary role here. Given its free fermion IR fixed point the only states available are formed from the perfectly free fermionic quasiparticles and according to textbook 
kinematics such states have vanishing energy in the long-wavelength limit. 
It follows immediately that, in the absence of this special kinematical protection, any strongly interacting IR fixed point of fermion 
matter (the non-Fermi liquid, by definition) will give rise to a plasmon that will in principle decay in the long-wavelength limit. 

{This can be seen explicitly in computations in holographic non-Fermi-liquid strange metals.}
In these holographic strange metals the other ``nonhydrodynamical'' charge carrying states are organized in a particularly elegant fashion, ruled by scaling principle,
 but the basic motive is completely general. {Do note that,} as we stressed in the Introduction, the quantum critical sector has a very different meaning in the holographic strange metal context
than in the case of quantum criticality originating in a quantum phase transition involving the disappearance of an order parameter. We expect, however,
that in these {latter} cases the plasmon will also decay. A case in point is the marginal Fermi liquid \cite{VarmaLittlewood,VarmaReview,2016RPPh...79h2501V}
and variations thereof. There a polarization propagator is envisioned which is not dissimilar
from the holographic one, but instead it originates in the quantum critical  fluctuations associated with the order parameter. 
It would be quite interesting to study the fate of the plasmon in this Herz-Millis framework, especially for the case of momentum conserving systems.

Zooming in on the particulars of the holographic plasmons, a quite surprising outcome is that the plasmon (as well as zero sound) is remarkably robust up to very 
large momenta. We find that in all cases we looked at, it is still underdamped up to momenta $|\bp| \sim \frac{1}{2}\mu$ --- (half) the Fermi momentum in a condensed matter interpretation. The plasmon at such momenta still dominates to a degree that the $\Xi(\omega,\bp)$ continuum is concealed from direct observation. In weakly interacting Fermi liquids the plasmon enters the Lindhard continuum at momenta {around this scale and is rapidly} obliterated by Landau damping. 
{There are basic scenarios where sound is less prominent, e.g. a generalization from static Coulomb interactions to relativistic electrodynamics case can make it more overdamped as we show in Appendix\,\ref{sec:Ax}. However, taking our lesson at heart, that the quantum critical sector is the generalization of the Lindhard continuum, the plasmon should be overdamped from the start. To understand why it is not is hard to answer.\footnote{Alternative approaches in triangular lattice Heisenberg antiferromagnets suggest that one possibility is that strong interactions may be responsible for taking the quasiparticles out of the  continuum \cite{Verresen2018}.} Why this ``hydrodynamic response'' works far better than it should --- an observation also made in other holographic studies ---} is a UV sensitive affair: the density operator of holography is associated with large central charge CFT physics, having \textit{a priori} nothing to do with the {microscopic electrons and atoms} of the condensed matter systems. Having the latter in mind, it is a matter of high priority to look for holographic set-ups that are more flexible in this regard.  A closely related issue is the incomplete understanding of how the quantum critical sector evolves at finite momenta.
{A starting point is \cite{Hartnoll:2012wm,Anantua2013,Gouteraux:2016arz},} but these are pressing issues for future holographic research. 

Last but not least, what is the relevance of this affair for the condensed matter laboratory?  We alluded to the timely nature of the subject in the Introduction, 
while presently major investments are made in condensed matter laboratories to improve the electron-loss instruments. To close the gap with experiment 
the highest priority is to do justice in holography to the circumstance that the non-Fermi-liquid electron systems in the laboratory are invariably characterized 
by very strong lattice potentials. Undoubtedly, the strong Umklapp scattering is a key factor in shaping the charge response.  Although this presents quite a 
technical challenge, this hurdle has to be taken in order to arrive at a meaningful  dialogue with experiment.  

\acknowledgments
The authors have benefited from discussions with 
Nikolay Gnezdilov,
Ulf Gran, Joerg Schmalian, Erik van Heumen and
Tobias Zingg.
This research was supported in part by a VICI award of the Netherlands Organization for Scientific Research (NWO), by the Netherlands Organization for Scientific Research/Ministry of Science and Education (NWO/OCW) and by the Foundation for Research into Fundamental Matter. We also wish to thank the organizers and participants of the workshop {\em Many-body Quantum Chaos, Bad Metals and Holography} made possible by support from NORDITA, ICAM (the Institute for Complex Adaptive Matter) and Vetenskapradet.

\begin{figure*}[ht]
	\hspace{-4mm}{\includegraphics[width=0.4\textwidth,scale=0.9]{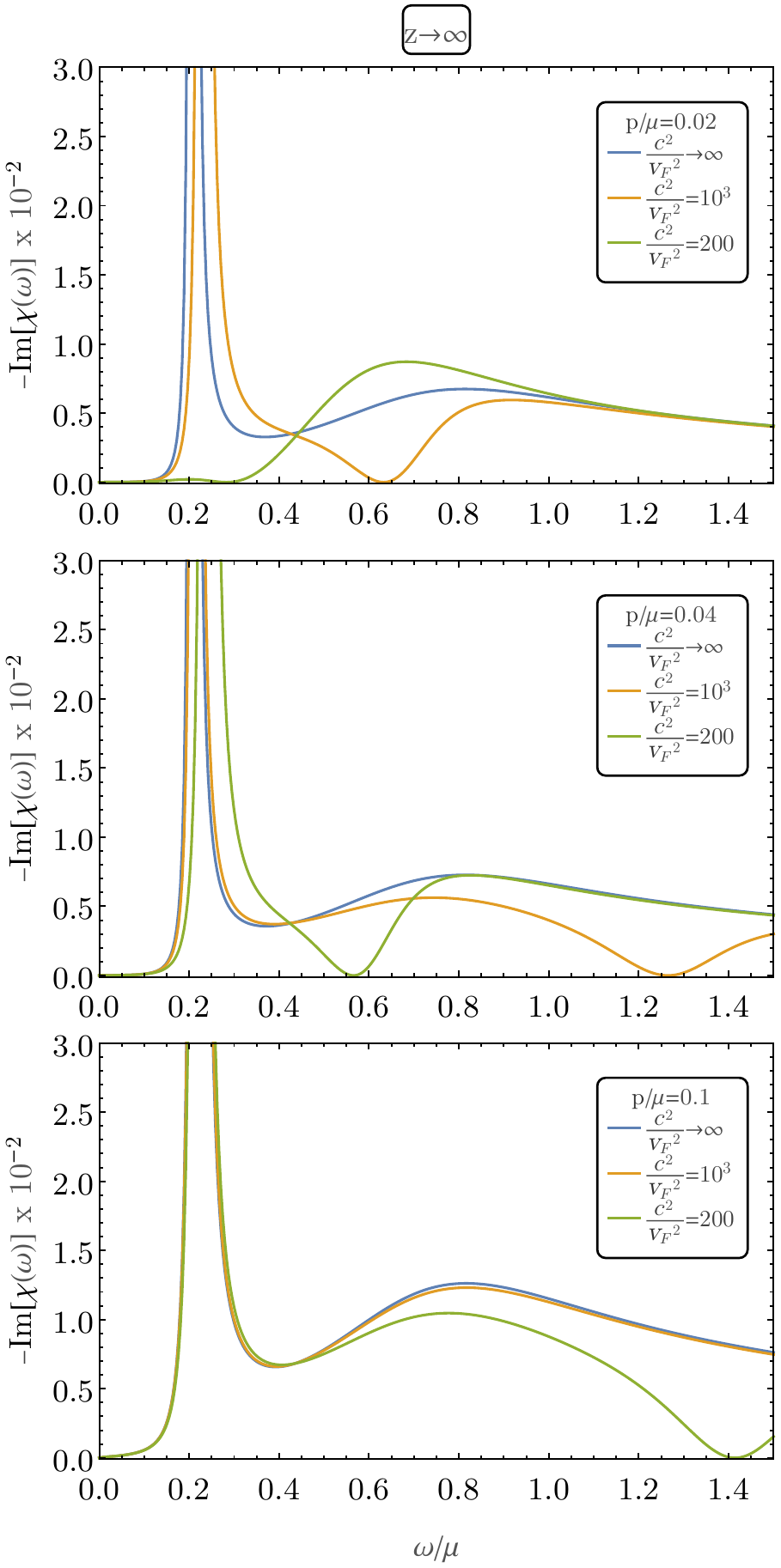}}
	~     \centering{\includegraphics[width=0.4\textwidth,scale=0.9]{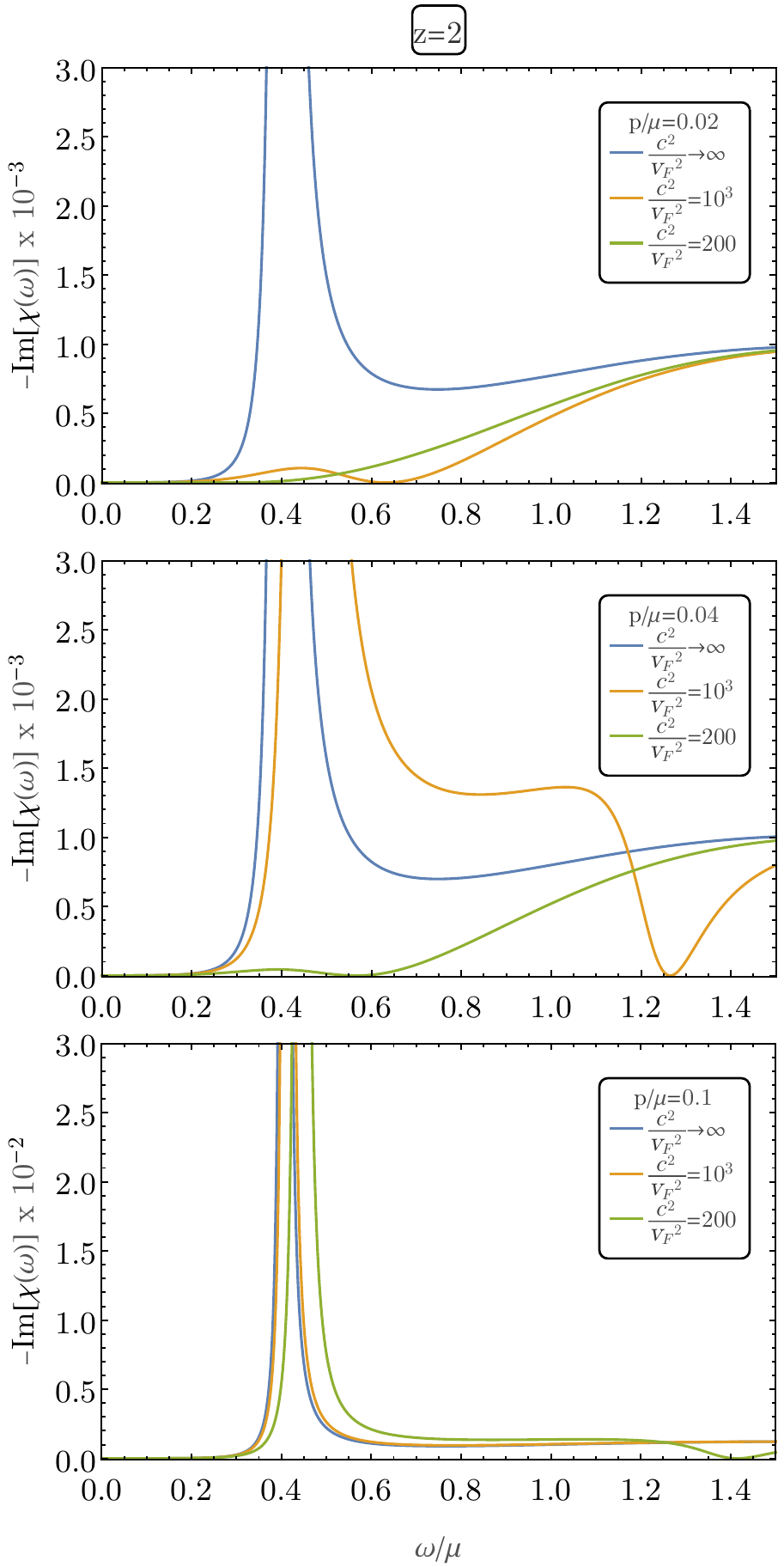}}
	\vspace{-5mm}
	\caption{Density-density response when both $V_{00}= -\frac{e^2 c^2}{c^2\bp^2-\omega^2}$ and $V_{xx}=\frac{e^2}{c^2\bp^2-\omega^2}$ are considered. Left column: $\theta=0,z\to\infty$, and right column $\theta=0,z=2$. The main differences observed with respect to the case of $V_{xx}\simeq0$, discussed in the main text, are the strong damping of the plasmon for larger momenta (green lines) and the zero of the spectral density, associated to the on-shell photon at $\omega=c|\bp|$.  We take the electromagnetic coupling $e^2=0.5$ and $T=0.02\mu$.}\label{fig:traceAx}
\end{figure*}

\appendix

\section{Gravitational equations of motion}
\label{app:EOMs}

The equations of motion for the background fields following from the action \eqref{equ:action} are \cite{Kiritsis:2015oxa}:
\begin{align}
R_{\mu\nu}-{2\Lambda\over d-1} g_{\mu\nu} &= {2\kappa\over q^2}Z(\phi)\left({1\over 2}F_{\mu\rho}F_{\nu}^{\phantom{\nu}\rho}-{F^2\over4(d-1)}g_{\mu\nu}\right)\notag\\
&+2\kappa\left(\half \p_\mu\phi\p_\nu\phi+{2\over d-1}V(\phi)\right)\notag\\
R_{\mu\nu}+3 g_{\mu\nu} &= {Z(\phi)\over2} \left(F_{\mu\rho}F_{\nu}^{\phantom{\nu}\rho}-{F^2\over4}g_{\mu\nu}\right)\notag\\
&+\half \p_\mu\phi\p_\nu\phi+{2\over d-1}V(\phi)\,,
\end{align}
where we set $2\kappa=q^2=1$.
The Maxwell and scalar equations are:
\begin{align}
\cov_\mu (Z(\phi)F^{\mu\nu})= 0\,,\  \cov^2\phi {-}W'(\phi){-}{Z'(\phi)\over 4}F^2=0\,.
\end{align}

These equations may be solved numerically using the shooting method in a particularly simple gauge \cite{Kiritsis:2015oxa,GarciaGarcia:2015}. However, here we choose to use the DeTurck method to solve the equations as a boundary value problem \cite{headrick2010new,adam2012numerical,horowitz2012optical, Rangamani:2015hka, Andrade:2017ghg}, which will facilitate our analysis of the linear fluctuations.
We use the metric ansatz 
for $d=3$:
\begin{align}
  \dd s^2 &= \frac{1}{u^2}\left( - Q_{tt}(u) f(u) \dd t^2 \right. \notag\\
  &\left.+ Q_{uu}(u) \frac{\dd u^2}{f(u)} + Q_{xx}(u) (\dd x^2 + \dd y^2) \right)\,, \label{equ:metric}\\
\notag
  f &= (1-u)\left( 1 + u + u^2 - \bar \mu^2 u^3 /4 \right)\,.
\end{align}

The DeTurck reference metric is just a Reissner-Nordstr\"om black hole which one gets when setting all the ansatz functions to unity. The temperature is set by the parameter $\bar{\mu}$ 
\begin{equation}\label{eq:T}
 	\frac{T}{\mu} = \frac{12 - \bar \mu^2}{16 \pi \mu},
\end{equation} 
and we can set $\bar{\mu} = \mu$ without loss of generality.
We consider $T/\mu=0.02$ throughout the paper, which is small enough and accessible for our numerics. 
We solve the resulting nonlinear elliptic equations of motion via Newton's method using the pseudospectral collocation approximation with 60 points on the pseudospectral grid. More details on the algorithm can be found in \cite{krikun2018numerical}.

\subsection{Linear response equations}

In order to evaluate the density-density correlator we first introduce the linear perturbations of all the fields with finite frequency and momentum:
\begin{equation}
\delta \phi \rar e^{- i \omega t + i p x} \delta \phi
\end{equation}
The analysis of the linearized equations shows that there are two decoupled composite modes:
\begin{align}
&\mathrm{longitudinal:} & &
\left.
\begin{cases}
\delta g_{tt}, \delta g_{tx}, \delta g_{tz}, \delta g_{xx}, \delta g_{xz},\\ \delta g_{yy}, \delta g_{zz},\delta A_t, \delta A_x, \delta A_z, \delta \phi   \label{equ:modes}
\end{cases}\hspace{-3mm}\right\}\\
&\mathrm{transverse:} & &\  \big\{ \delta g_{ty}, \delta g_{xy}, \delta g_{yz},  \delta A_y\big\} 
\end{align}
We use the DeDonder gauge in $\delta g_{\mu \nu}$ and Lorentz gauge in $\delta A_\mu$ \cite{horowitz2012optical, Rangamani:2015hka}. This brings all the equations of motion for the 11 fields in the longitudinal channel to the convenient second order form. 

Using the pseudospectral collocation approximation on the same grid as the background solutions, we recast the linearized equations as a linear algebraic problem
\begin{equation}
\mathcal{M} \vec{f} = \vec{\cal{A}},
\end{equation}
where the right hand side corresponds to the source in \eqref{equ:dirichlet}. This equation is efficiently solved by the \texttt{LinearSolve} routine of Wolfram Mathematica \cite{Mathematica}, giving the vector representation of the linearized solutions, which we use to read off the two-point function \eqref{equ:b_over_a}. 

The quasinormal modes are the solutions to the Sturm-Liuville problem 
\begin{equation}
\mathcal{M} \vec{f} = 0,
\end{equation} 
and are simply the eigenvalues of the matrix $\mathcal{M}(\omega)$ (This is similar to the method used in \cite{andrade2018pinning}). 

We also get the plasmon modes in a similar way, but one should note that in this case we use the different boundary conditions \eqref{eq:Robin}, and therefore the matrix of the boundary value problem gets modified \cite{krikun2018numerical}. The plasmon modes are therefore the eigenvalues of the deformed matrix $\mathcal{M}_{V_\bp}(\omega)$.

\begin{figure*}[t]
	\hspace{-5mm}
	{\includegraphics[scale=0.53]{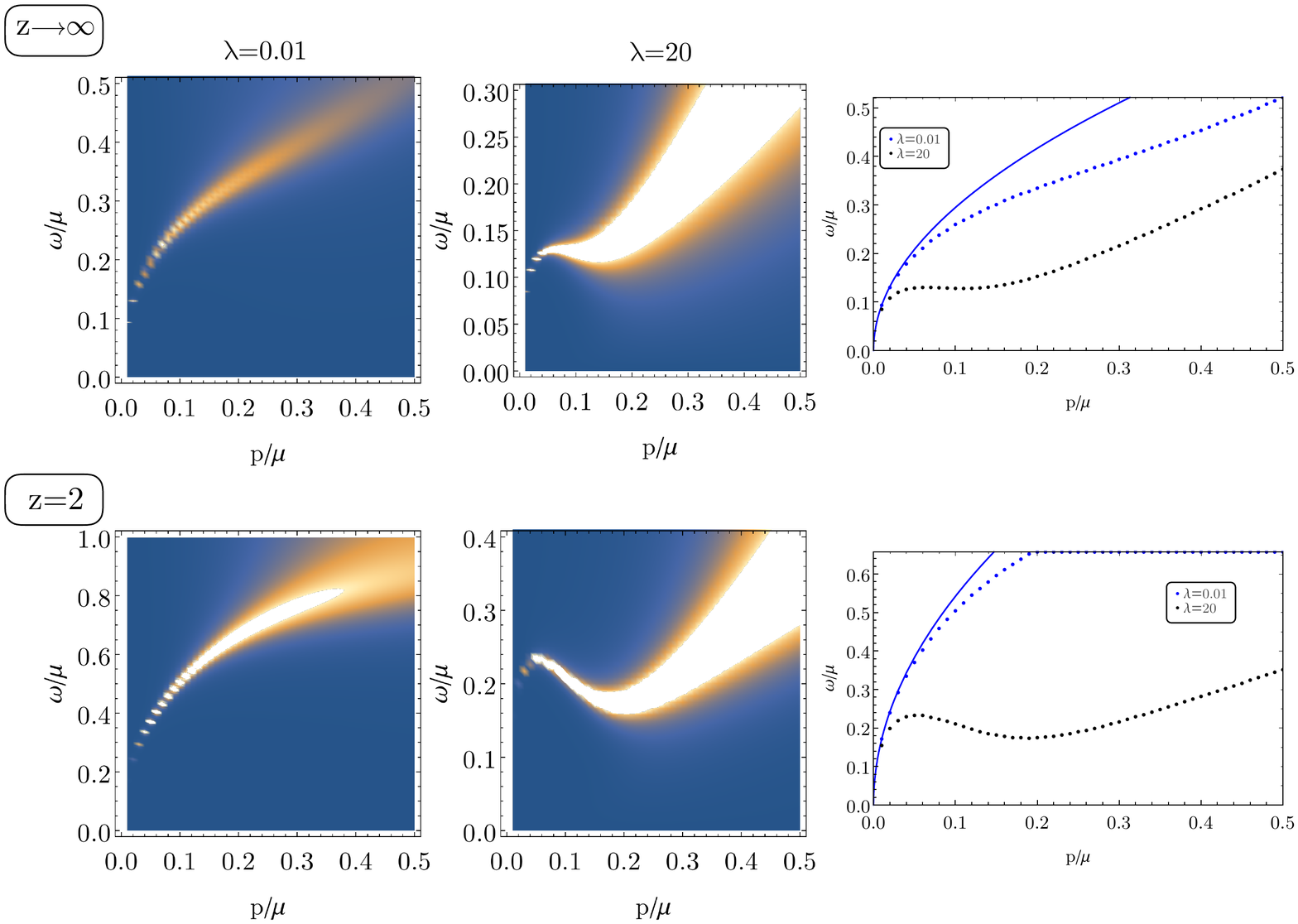}}
	\vspace{-3mm}
	\caption{Dressed density-density response $-\Im\chi$ from Eq.\,\eqref{eq:chi_dressed} with $V_{\bp}=e^{-\lambda |\bp|}/|\bp|$ for  $\theta=0$ and $z\to\infty$ (top row) and $z=2$ (bottom row). The parameter space of the effective Coulomb interaction in layered systems is qualitatively captured by the two values presented: $\lambda=0.01\mu$ (left column) and $\lambda=20\mu$ (middle column). On the right column we show the  simplified dispersion relations with only the maxima drawn as dots, and square-root lines as a guide to the eye.}\label{fig:exponential}
\end{figure*}

\section{From Coulomb interaction to full dynamical electromagnetism}
\label{sec:Ax} 

The nonrelativistic approximation and reduction of full dynamical electromagnetism to the static Coulomb potential only [Eq. \eqref{eq:2}] clearly breaks Lorentz invariance. This is justified because the Lorentz invariance, implicit in the general relativistic description of holographic strange metals, is set by the emergent speed $v_F$ in the linear dispersion relation of the gapless modes at low energies and not the fundamental speed of light in vacuum $c$. Nevertheless, it is straightforward to include dynamical electromagnetism by either the proper photon propagator $V_{\mu \nu} = i D_{\mu \nu} = \eta_{\mu \nu}/p^2$ or a generalized one that takes into account the difference between $v_F=1$ and $c$: $V_{00} = -e^2 c^2/(c^2\bp^2-\omega^2), V_{xx} = \frac{e^2}{c^2\bp^2-\omega^2}$. The full electromagnetic interaction $J^\mu V_{\mu \nu} J^\nu$ will couple both charge densities $n \equiv J_0$ via electric Coulomb force, but also the spatial components of the currents $J_i$ via the magnetic Lorentz force. The RPA formula \eqref{eq:RPA} becomes matrix valued and gives
\onecolumngrid
\begin{equation}
\label{eq:full_response}
\hspace*{-.2in}
\begin{bmatrix}
    \chi_{00}       & \chi_{0x}\\
    \chi_{x0}       & \chi_{xx} \\
\end{bmatrix}
={1\over 1+V_{00}\Pi_{00}+V_{xx}\Pi_{xx}-\textrm{det}\Pi \ V_{00}V_{xx}}
\begin{bmatrix}
    \Pi_{00}+V_{xx}\ \textrm{det}\Pi &\Pi_{0x} \\
    \Pi_{x0}&   \Pi_{xx}+V_{00}\ \textrm{det}\Pi 
\end{bmatrix},
\end{equation}
\twocolumngrid
\noindent where $\Pi_{\mu\nu} = \langle J_{\mu}J_{\nu}\rangle_{e^2=0}$ and $\textrm{det}\Pi=(\Pi_{00}\Pi_{xx}-\Pi_{x0}\Pi_{0x})$. 
In the nonrelativistic limit $c\rightarrow \infty$ this reduces to the previous results as should be.

For completeness we give an example of the plasmon response with dynamical electromagnetism with $c^2/v_F^2 = 10^3,200$ and $e^2=0.5$ in Fig.\,\ref{fig:traceAx}. At $\bp \to0$, $V_{xx}$  has a similar effect to a stronger $V_{00}$ dressing, namely the gap increases and the plasmon peak is damped more significantly. There is however, a qualitatively different feature: the on-shell photon state, which is reflected as a zero of the density-density response at $\omega= c|\bp|$. 
As momentum increases there is another qualitative difference:  the gap and peak damping initially decrease as momentum increases until the effect of $V_{xx}$ is negligible and the response is controlled again by $V_{00}$ alone.

\section{\label{app:dressings} 2D Coulomb interaction}

\begin{figure}[h!]
\centering
    \centering{\includegraphics[scale=0.8]{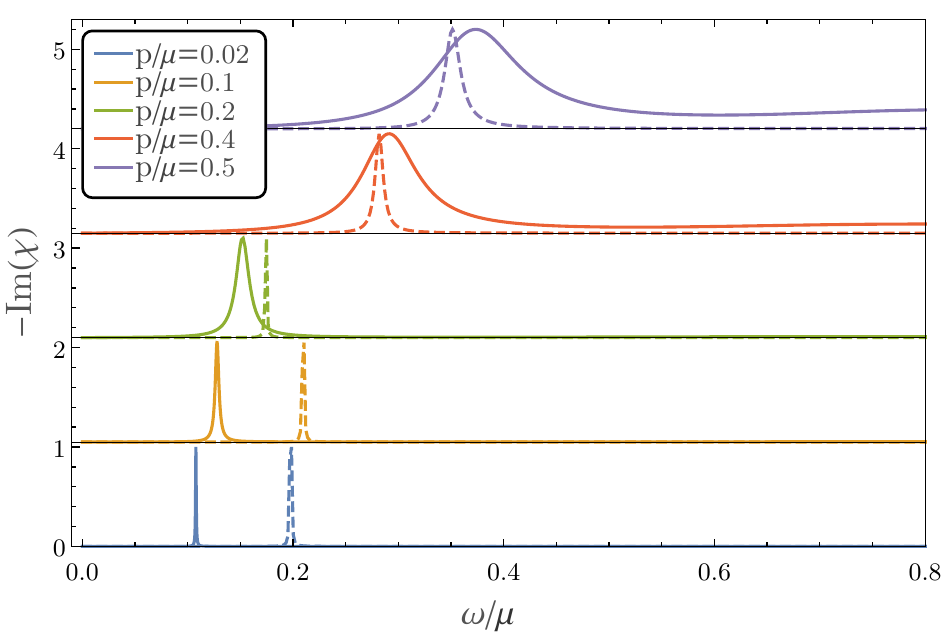}}
    \vspace{-3mm}
  \caption{Momentum slices of the density-density response $-\Im\chi(\omega,\bp)$ shown in Fig.\,\ref{fig:exponential}. The continuous lines correspond to  $\theta=0$, $z\to\infty$ and the dashed lines to $z=2$, $\theta=0$. This effective Coulomb interaction found in layered systems makes the collective excitation narrow at small momenta.}\label{fig:slices_exp}
\end{figure}

In layered systems the Coulomb interaction is effectively two dimensional: $V(|\bp|)=e^{-\lambda |\bp|}/|\bp|$. This interaction leads to  the standard square-root dispersion relation (see Fig.\,\ref{fig:exponential}). Moreover, the effects of the quantum critical sector on the plasmon width are hardly visible since for low momentum $\bp\to0$ (see Fig.\,\ref{fig:slices_exp}), because  the width of the plasmon peak vanishes  as
$\tilde\Gamma= \Gamma + AV_{\bp}^2\bp^4 (-\text{Im}(\Xi)/\omega)+\ldots\sim \OO(|\bp|^2)$. 

\newpage
\bibliography{plasmon}

\end{document}